\documentclass[%
 reprint,
superscriptaddress,
 amsmath,amssymb,
 aps,
pre,
]{revtex4-2}

\usepackage{xcolor}
\usepackage[normalem]{ulem}

\usepackage{graphicx}
\usepackage{dcolumn}
\usepackage{bm}
\usepackage{texshade}
\usepackage{amsmath}
\usepackage{amssymb}
\usepackage{mathtools}
\usepackage{float}
\usepackage{soul}
\usepackage{tikz}
\usepackage[title]{appendix}

\usepackage{appendix}

\usetikzlibrary{shapes.geometric, arrows}

\definecolor{cbBlue}{HTML}{0072B2}
\definecolor{cbOrange}{HTML}{E69F00}
\definecolor{cbGreen}{HTML}{009E73}
\definecolor{cbVermillion}{HTML}{D55E00}
\definecolor{cbPurple}{HTML}{CC79A7}
\definecolor{esNavy}{HTML}{1B4965}
\definecolor{esOlive}{HTML}{6A994E}
\definecolor{esSand}{HTML}{D4A373}
\definecolor{esCrimson}{HTML}{A63D40}
\definecolor{dustyBlue}{HTML}{5E7896}
\definecolor{mutedTeal}{HTML}{6FA7A1}
\definecolor{mauve}{HTML}{9A6C87}
\definecolor{slateGray}{HTML}{5F636E}

\tikzstyle{startstop} = [rectangle, rounded corners, minimum width=0.15\columnwidth,
    minimum height=0.8cm, text centered, draw=black, fill=red!20,
    text width=0.43\columnwidth, align=center, font=\small]
\tikzstyle{process} = [rectangle, minimum width=0.15\columnwidth, minimum height=0.8cm,
    text centered, draw=black, fill=blue!20,
    text width=0.43\columnwidth, align=center, font=\small]
\tikzstyle{decision} = [diamond, aspect=2, draw=black, fill=green!20,
    text width=0.36\columnwidth, align=center, inner sep=0pt, font=\small]
\tikzstyle{arrow} = [thick,->,>=stealth]

\begin{document}


\title{A route to the thermodynamics of colloid-polymer mixtures from structural information}

\author{Vikki Anand Varma}
\affiliation{Department of Materials \& Mechanical Engineering, University of Turku, Turku 20500, Finland}

\author{Andrew J. Archer}
\affiliation{Department of Mathematical Sciences \& Interdisciplinary Centre for Mathematical Modelling, Loughborough University, Loughborough LE11 3TU, United Kingdom}

\author{Alberto Scacchi}
\email{alberto.scacchi@utu.fi}
\affiliation{Department of Materials \& Mechanical Engineering, University of Turku, Turku 20500, Finland}

\date{\today}

\begin{abstract} 
Liquid-state theory provides a fundamental connection between microscopic structure and macroscopic thermodynamic behaviour. Here, we develop a framework for predicting thermodynamics and phase behaviour directly from structural correlations, using either radial distribution functions or static structure factors as input. The approach constructs a free-energy functional from structural information obtained at a single thermodynamic state point, without requiring explicit knowledge of the underlying interaction potentials. This circumvents a central difficulty in modelling complex fluids, for which effective interactions are often unknown or rely on approximations. We demonstrate the framework for a range of model fluids, including colloidal and colloid--polymer systems, with predictions in good agreement with molecular simulation data. The results show that structural information at a single state point can provide sufficient information to predict the broader thermodynamic response of a system. This establishes a route toward inferring phase behaviour directly from experimentally measured structure, even when the microscopic interactions are not known {\it a priori}.
\end{abstract}

\maketitle

\section{Introduction}

\noindent Colloid–polymer mixtures are prototypical soft-matter systems and are ubiquitous in both industrial formulations and biological materials, where they contribute to the organization of complex living structures \cite{RevModPhys.64.645, Hamley2007}. Owing to their tunable effective interactions and rich phase behaviour \cite{poon1994phase, WCKPoon_2002, likos2001effective}, they have enabled a wide range of applications in biotechnology and environmental engineering, including drug delivery, biomolecule purification, and water treatment via aqueous two-phase extraction \cite{Fick2023, Zhang2024}. Among the diverse phenomena exhibited by these systems, liquid–liquid phase separation of polymers in the presence of colloids is of particular interest \cite{wennerstrom2026colloidal}. The resulting partitioning of colloids between coexisting polymer-rich phases underlies separation processes and has also been implicated in intracellular compartmentalization and other biological phase-separated environments \cite{banani2017biomolecular}. Understanding these systems therefore requires an accurate description of their thermodynamics, including phase stability, colloidal partitioning, osmotic pressure, and interfacial properties such as wetting and surface tension \cite{brader2003statistical,nilson2006influence,llovell2010classical}.

Over the past several decades, liquid-state theory has established a rigorous framework for relating the microscopic structure of fluids to their macroscopic thermodynamic properties. Central to this description are the radial distribution function (RDF), $g(r)$, the static structure factor, $S(k)$, and the pair direct correlation function (DCF), $c^{(2)}(r)$, which characterize equilibrium fluid structure and correlations \cite{hansen2013theory}.
Note that we are assuming here that interactions are radially symmetric, so $g(r)$ is only a function of $r$, the distance between pairs of particles.
Thermodynamic properties can be obtained from these quantities through well-established routes such as the virial and compressibility equations, while the DCF plays a central role in classical density functional theory (cDFT) \cite{evans1979nature, evans1992density}, providing the link between microscopic correlations and free-energy functionals \cite{evans1978long, ramakrishnan1979first, lutsko2007density}. Consequently, cDFT has become one of the principal theoretical frameworks for predicting phase equilibria and interfacial properties of inhomogeneous fluids.
In practice, however, obtaining accurate thermodynamic properties generally requires evaluating structural correlations over a range of thermodynamic state points together with thermodynamic integration (TI), making such calculations computationally demanding \cite{hansen2013theory}. An alternative strategy is to reconstruct effective pair interaction potentials from structural information and subsequently determine the thermodynamics from the resulting interactions. At low densities, the potential of mean force, $\phi(r)=-k_{\rm B}T\ln g(r)$, provides an estimate of the pair interaction, but this approximation neglects many-body correlations and therefore breaks down at finite densities \cite{hansen2013theory}. More sophisticated inverse methods, such as inverse Boltzmann iteration (IBI), recover effective interaction potentials from RDFs at finite density \cite{moore2014derivation}; however, the resulting potentials generally retain a dependence on the thermodynamic state point.

These considerations motivate the inverse framework developed in the present work, in which thermodynamic properties are obtained directly from structural correlations rather than reconstructed interaction potentials. The approach requires only either the RDF, or equivalently the static structure factor, at a single thermodynamic state point. We present two complementary routes. The first determines the thermodynamics directly from the RDF, while the second is based on the DCF, which is obtained from the structure factor through the Ornstein--Zernike (OZ) relation~\cite{hansen2013theory}, $S(k)=[1-\rho\hat{c}(k)]^{-1}$, where $\hat{c}(k)$ denotes the Fourier transform of the DCF. The resulting free-energy functional can then be employed within cDFT to predict phase behaviour and interfacial properties without the extensive TIs required by conventional approaches or explicit knowledge of the underlying interaction potentials.

To demonstrate the generality of the framework, we first apply it different one-component model colloidal fluids before extending it to colloid--polymer mixtures. In the latter case, we combine a DCF-based treatment of attractive interactions with an effective description of excluded-volume effects arising from the colloids, yielding a consistent free-energy functional for mixtures with interacting polymers. We validate the approach through comparison with both newly generated and existing simulation data, before applying it to predict the phase behaviour of a specific colloid--polymer mixture.

\section{Theory}
\label{sec:II}

The present formalism is developed for mixtures; however, for clarity we first consider a one-component system.
The goal is to develop a framework for determining the thermodynamic properties of a system, including its bulk phase behaviour, from $g(r)$ (or, equivalently, $S(k)$) evaluated at a single thermodynamic state points.

Given $g(r)$, we distinguish two broad classes of interactions. In systems with purely soft interactions, such as polymeric systems where the soft `particles' can interpenetrate, $g(r)$ remains finite over all relevant length scales \cite{likos2001effective}. In contrast, systems with excluded-volume interactions exhibit a finite distance $\sigma_{\rm u}$ such that
$g(r)=0$ for $r<\sigma_{\rm u}$, where $\sigma_{\rm u}$ defines an effective core diameter.
In these cases, the interaction potential is assumed to have a decomposable form given as,
\begin{equation}
\phi(r)=\phi^{\rm ref}(r)+\phi^{\rm tail}(r),
\label{eq::potential_decomposition}
\end{equation}
where $\phi^{\rm ref}(r)$ describes the short-range repulsive core, while $\phi^{\rm tail}(r)$ represents the remaining (tail) contribution (attractive or repulsive) which remains finite at all separations.
We assume that the contribution from the core can be mapped onto a hard-sphere reference system with an optimal diameter $\sigma_{\rm opt}\neq\sigma_{\rm u}$~\cite{percus1958analysis, wertheim1964analytic, mansoori1971equilibrium, hansen2013theory}. A novelty of our approach is the way we determine $\sigma_{\rm opt}$.
In the case of a known form of $\phi(r)$, the typical way to estimate the effective hard-sphere diameter is using the Barker-Henderson (BH) approach, namely~\cite{barker1967perturbation}
\begin{equation}
\sigma_{\rm BH}
=
\int_0^{\infty}
\left[
1-e^{-\beta \phi^{\rm ref}(r)}
\right]dr,
\label{eq::sigma_bh}
\end{equation}
where $\beta=(k_BT)^{-1}$, with $k_{\rm B}$ the Boltzmann factor and $T$ the temperature, and where $\phi^{\rm ref}$ can be obtained following various schemes. For example, $\phi^{\rm ref}$ can be defined using the Weeks-Chandler-Anderson (WCA) splitting~\cite{weeks1971role}, in which the potential is truncated at a distance corresponding to its minimum and shifted so that it vanishes at that point. However, here we assume that the pair potential, $\phi(r)$, is unknown and therefore adopt a different approach.

\subsection{Effective hard-sphere diameter}
\label{sec:IIA}

The simplest way to estimate an effective diameter directly from a RDF is from the largest distance at which $g(r)=0$, yielding $\sigma_{\rm u}$. A better estimate can be obtained by first recovering an effective interaction potential from the RDF using the IBI method (see details in Sec.~\ref{sec:A1} of the Appendix). The effective IBI potential can be split using the WCA approach, providing an estimate for the reference potential, $\phi^{\rm ref}(r)$. Then, the OZ equation, together with an appropriate closure relation (see the Appendix, Secs.~\ref{sec:A2} and \ref{sec:A3}), here the Percus-Yevick (PY) closure~\cite{PhysRev.110.1,pihlajamaa2024comparison}, is solved to obtain the corresponding correlation function $g^{\mathrm{ref}}(r)$. $\sigma_{\rm opt}$ is then determined by minimizing
\begin{equation}
\mathcal{D}(\sigma_{\rm opt})
=
\int_{0}^{r_{\mathrm{c}}}
\left[
g^{\mathrm{ref}}(r)
-
g^{\mathrm{HS}}(r;\sigma_{\rm opt})
\right]^2
dr,
\label{eq:diameter_fitting}
\end{equation}
where $g^{\mathrm{HS}}(r;\sigma_{\rm opt})$ denotes the hard-sphere RDF corresponding to $\sigma_{\rm opt}$.
Noise is often present in a RDF measured either experimentally or from simulations. To minimize the influence of such noise, we introduce a cutoff distance, $r_{\mathrm{c}}'$, based on the positions of the peaks in $g(r)$. Specifically, we determine the average separation between successive peaks, denoted by $r_{\rm p}$. For distances greater than $r_{\rm last}+r_{\rm p}$, where $r_{\rm last}$ is the position of the last identified peak, the measured RDF is set to unity beyond $r_{\mathrm{c}}'=r_{\rm last}+r_{\rm p}$, i.e., $g(r)=1$ for $r>r_{\rm c}'$. In systems exhibiting only a single peak, which typically occurs at low densities, the cutoff distance is taken as $r_{\mathrm{c}}'=2r_{\rm p}$. 

Multi-component systems, for which the same idea applies, are discussed in more detail in the Appendix; see also Refs.~\cite{hanke2018well,moore2014derivation}.

For systems interacting purely through soft potentials, the excluded-volume contribution vanishes and no effective size estimation is required. In such cases, the thermodynamics can be computed directly from the RDF and interaction potential as detailed below.

\subsection{Helmholtz free energy}

To extract the thermodynamics of a system purely from structural data, we begin by expressing the total Helmholtz free energy as~\cite{evans1979nature, hansen2013theory}
\begin{equation}
F[\rho(\mathbf{r})] =
F^{\rm id}[\rho(\mathbf{r})] + F^{\rm exc}[\rho(\mathbf{r})],
\label{eq:landau_potential}
\end{equation}
where the first term represents the exact ideal-gas contribution, $\beta F^{\rm id}[\rho] = \int_{V} d\mathbf{r}\,\rho(\mathbf{r}) \left[\ln(\Lambda^3\rho(\mathbf{r})) - 1\right]$, where $\rho$ is the one-body number density, $\Lambda$ is the de Broglie wavelength, and $V$ the volume of the system. $F^{\rm exc}[\rho(\mathbf{r})]$ denotes the excess free energy function which is the contribution due to particle interactions via the pair potential $\phi$, which can be decomposed as in Eq.~(\ref{eq::potential_decomposition}). In the same way, the excess free energy functional can be further split into\begin{equation}
F^{\rm exc}[\rho(\mathbf{r})] =
F^{\rm exc}_{\rm ref}[\rho(\mathbf{r})] +
F^{\rm exc}_{\rm tail}[\rho(\mathbf{r})].
\label{eq:free_energy_decomposition}
\end{equation}
Here, $F^{\rm exc}_{\rm ref}$ defines the reference-system free energy arising from excluded volume effects through $\phi^{\rm ref}$, while $F^{\rm exc}_{\rm tail}$ describes the remaining part (attractive or repulsive) of the interactions, treated perturbatively through $\phi^{\rm tail}$. By mapping $\phi^{\rm ref}$ onto an effective hard-sphere system as described above, $F^{\rm exc}_{\rm ref}$ can be approximated by the excess free energy of the corresponding hard-sphere fluid.
For the bulk fluid, the PY approximation \cite{percus1958analysis, wertheim1964analytic} or even better, the Carnahan-Starling equation of state \cite{mansoori1971equilibrium, hansen2006new}, provide suitable approximations. Here, we develop the formalism to include inhomogeneous fluids, so $F^{\rm exc}_{\rm ref}$ may therefore be evaluated using fundamental measure theory (FMT), which in bulk generates either the PY or Carnahan-Starling equation of state, depending on the particular implementation \cite{rosenfeld1989free, Roth_2010}.

The remaining contribution to be determined, $F_{\rm tail}^{\rm exc}$, is described in the next section.

\subsection{Excess free energy}

A widely used approach for obtaining the free energy contribution from the tail is via TI \cite{hansen2013theory}.
The often used starting point is the result that the functional derivative of the free energy with respect to the pair potential $\phi(r)$ is related to the two-body density distribution function $\rho^{(2)}(\mathbf{r},\mathbf{r}')$, as~\cite{evans1979nature}
\begin{equation}\label{eq:rho2}
    \rho^{(2)}(\mathbf{r},\mathbf{r}')=2\frac{\delta F[\rho]}{\delta \phi(\mathbf{r},\mathbf{r}')}.
\end{equation}
Starting from the free energy of a reference system with pair potential $\phi^\mathrm{ref}(r)$ and `turning on' the tail of the potential via the coupling parameter $\alpha\in[0,1]$, so that $\phi_\alpha(r)=\phi^\mathrm{ref}(r)+\alpha\phi^\mathrm{tail}(r)$, yields a following formally exact result for $F^{\mathrm{exc}}_{\rm tail}[\rho(\mathbf{r})]$, namely~\cite{evans1979nature, 10.1063/1.4993175}
\begin{equation}
F^{\mathrm{exc}}_{\rm tail}[\rho(\mathbf{r})]
=
\frac{1}{2}
\int_0^1 d\alpha
\int d\mathbf{r}\int d\mathbf{r}'
\rho^{(2)}(\mathbf{r},\mathbf{r}';\alpha)
\phi^{\rm tail}(|\mathbf{r}-\mathbf{r}'|),
\label{eq::exact_free_energy)_att}
\end{equation}
where $\rho^{(2)}(\mathbf{r},\mathbf{r}';\alpha)$ denotes the two-body density at coupling parameter $\alpha$.
Writing $\rho^{(2)}(\mathbf{r},\mathbf{r}';\alpha) =\rho(\mathbf{r}) \rho(\mathbf{r}') g_\alpha(\mathbf{r},\mathbf{r}')$ \cite{hansen2013theory}, where $g_\alpha$ is the RDF for the system at $\alpha$, the attractive contribution becomes
\begin{equation}
F^{\mathrm{exc}}_{\rm tail}[\rho(\mathbf{r})]
=
\frac{1}{2}
\int d\mathbf{r}\int d\mathbf{r}'
\rho(\mathbf{r})
\rho(\mathbf{r}')
G(\mathbf{r},\mathbf{r}')
\phi^{\rm tail}(|\mathbf{r}-\mathbf{r}'|),
\label{eq:thermodynamic_integration}
\end{equation}
where
\begin{equation}
G(\mathbf{r},\mathbf{r}')
=
\int_0^1 d\alpha\,
g_\alpha(\mathbf{r},\mathbf{r}')
\label{eq::g_alpha}
\end{equation}
is the coupling-averaged pair correlation function.

For a homogeneous fluid, where $\rho(\mathbf{r})=\rho_0$, we have $\rho^{(2)}(\mathbf{r},\mathbf{r}';\alpha)=\rho_0^2 g_\alpha(r)$, where $r=|\mathbf{r}-\mathbf{r}'|$, yielding
\begin{equation}
\frac{F^{\mathrm{exc}}[\rho_0]}{V}
=
\frac{F^{\mathrm{exc}}_{\mathrm{ref}}[\rho_0]}{V}
+
\frac{1}{2}\rho_0^2
\int_0^1 d\alpha
\int d\mathbf{r}\,
g_\alpha(r)\,
\phi^{\rm tail}(r),
\label{eq::exact_free_energy_gr}
\end{equation}
where the last term corresponds to $F^{\rm exc}_{\rm tail}[\rho_0]/V$.

We now define the integrated interaction strength,
\begin{equation}
A_{\rm TI}
=
4\pi
\int_0^\infty
r^2
G(r)\,
\phi^{\rm tail}(r)\,
dr
=
4\pi
\int_0^\infty
r^2
a_{\rm TI}(r)\,
dr,
\label{eq::integrated_strength_gr}
\end{equation}
where $a_{\rm TI}(r)=G(r)\phi^{\rm tail}(r)$ is the integration kernel.
Using Eq.~(\ref{eq::integrated_strength_gr}), the excess free energy for the bulk system can be written as
\begin{equation}
\frac{F^{\mathrm{exc}}[\rho_0]}{V}
=
\frac{F^{\mathrm{exc}}_{\mathrm{ref}}[\rho_0]}{V}
+
\frac{1}{2}\rho_0^2 A_\mathrm{TI}.
\label{eq::exact_free_energy_final}
\end{equation}
Here, for simplicity, we approximate $a_{\rm TI}$, and hence the corresponding integrated strength $A_{\rm TI}$, using
\begin{equation}
a_{\rm TI}(r)=g(r)\phi^{\rm tail}(r),
\end{equation}
where $\phi^{\rm tail}(r)$ is the attractive tail obtained from the IBI procedure. This approximation assumes that $g_{\alpha}(r)$ remains approximately unchanged as the attractive tail is gradually introduced, such that $g_{\alpha}(r)\simeq g_{\alpha=1}(r)$ over the integration path. This is motivated by the assumption that the structural correlations are dominated by the excluded-volume interactions, while the attractive tail provides a comparatively weaker contribution to the structure. Consequently, the full-system RDF, $g_{\alpha=1}(r)$, can be used to approximate the structural correlations throughout the integration, eliminating the need to evaluate $g_{\alpha}(r)$ at multiple values of $\alpha$ (which cannot be done experimentally).

\subsection{Excess free energy via $c^{(2)}(r)$}

Instead of starting from Eq.~\eqref{eq:rho2} and integrating to obtain the free energy in Eq.~\eqref{eq::exact_free_energy)_att}, an alternative approach is to start from the following result: the pair direct correlation function is the second derivative of the free energy with respect to the one-body density, namely~\cite{evans1979nature, hansen2013theory}
\begin{equation}
    \label{eq:c2}
    c^{(2)}(\mathbf{r},\mathbf{r}')=-\beta\frac{\delta^2 F[\rho]}{\delta \rho(\mathbf{r})\delta\rho(\mathbf{r}')}.
\end{equation}
Integrating this twice, we obtain the following for the excess free energy \cite{evans1979nature}
\begin{equation}
\begin{split}
\beta F^{\mathrm{exc}}[\rho(\mathbf{r})]
&=
\int_0^1 d\lambda  (\lambda - 1) \\&
\times \int d\mathbf{r}\int d\mathbf{r}'
\rho(\mathbf{r})\rho(\mathbf{r}')
c^{(2)}_\lambda(\mathbf{r},\mathbf{r}'),
\end{split}
\label{eq:exact_free_energy_v2}
\end{equation}
where the integration parameter $\lambda\in[0,1]$, and the integration is through a series of states with density profile $\lambda \rho(\mathbf{r})$. The function $c^{(2)}_\lambda(\mathbf{r},\mathbf{r}')$ is the corresponding DCF along the integration path.
Splitting $c^{(2)}_\lambda$ into a part corresponding to a reference system (excluded volume contribution), $c^{(2)}_\mathrm{ref}$, and a remainder (tail contribution), this gives
\begin{eqnarray}\label{eq:exact_free_energy_v2_split}
F^{\mathrm{exc}}[\rho(\mathbf{r})]
&=&F^{\mathrm{exc}}_\mathrm{ref}[\rho(\mathbf{r})]\\
&\,&
\hspace{-1.2cm}+k_{\rm B}T
\int_0^1  d\lambda (\lambda-1)
\int d\mathbf{r}\int d\mathbf{r}'
\rho(\mathbf{r})\rho(\mathbf{r}')
\Delta c^{(2)}_\lambda(\mathbf{r},\mathbf{r}'),
\nonumber
\end{eqnarray}
where $\Delta c^{(2)}_\lambda=c^{(2)}_\lambda-c^{(2)}_\mathrm{\lambda, ref}$.
In bulk this gives \cite{evans1992density}
\begin{equation}
    \frac{F^{\mathrm{exc}}[\rho_0]}{V}
=
\frac{F^{\mathrm{exc}}_{\mathrm{ref}}[\rho_0]}{V}
+
k_{\rm B}T\rho_0^2
\int_0^1  d\lambda (\lambda - 1)
\int d\mathbf{r}\,
\Delta c^{(2)}_\lambda(r).
\label{eq:bulk_free_c2}
\end{equation}
By assuming $\int_0^1  d\lambda (\lambda - 1 )  \Delta c^{(2)}_\lambda(r)\approx -\frac12\Delta c^{(2)}(r)$, we get
\begin{equation}
    \frac{F^{\mathrm{exc}}[\rho_0]}{V}
=
\frac{F^{\mathrm{exc}}_{\mathrm{ref}}[\rho_0]}{V}
-\frac12 k_BT\rho_0^2
\int d\mathbf{r}\,
\Delta c^{(2)}(r),
\label{eq:bulk_free_c2_approx}
\end{equation}
where in this approximation we effectively assume that the value of $\Delta c_\lambda^{(2)}(r)$ is independent of $\lambda$ and equal to $\Delta c^{(2)}(r)$, the value at $\lambda  = 1$.

Comparing Eqs.~(\ref{eq::exact_free_energy_gr}) and (\ref{eq:bulk_free_c2_approx}), it is tempting to identify $G(r)\phi_{\rm tail}(r)$ with $-k_{\rm B}T\Delta c^{(2)}(r)$. However, this identification is not generally valid, as the two expressions arise from different approximations and therefore need not be equivalent.
However, in analogy with Eq.~(\ref{eq::integrated_strength_gr}), we do denote the integral in \eqref{eq:bulk_free_c2_approx} as
\begin{equation}
A_{c^{(2)}}
=
- k_{\rm B}T\, 4\pi \int r^2\, \Delta c^{(2)}(r)\,dr,
\label{eq::integrated_strength_delta_cr}
\end{equation}
where $A_{c^{(2)}}$ is the integrated attraction strength obtained via the DCF route. Here, the integrated attraction strength kernel is given as $a_{c^{(2)}} = - k_{\rm B}T \Delta c^{(2)}(r)$.

Here, we compute $c^{(2)}_{\rm ref}$ for the reference system via the OZ equation (see Appendix) together with the PY closure, where the reference system is approximated by an effective hard-sphere fluid with diameter $\sigma_\mathrm{opt}$, following the procedure described in Sec.~\ref{sec:IIA}.

\subsection{Polymer free energy}

As mentioned at the beginning of this section, if $g(r)>0 \ \forall r$, there is no reason to define $F^{\rm exc}_{\rm ref}$. This feature of the RDF can be seen in several models for, e.g., linear polymers and dendrimers~\cite{Louis20002522,Bolhuis20014296, likos2001effective, Gotze20047761,10.1063/1.2172596, PhysRevE.64.041501, PhysRevResearch.3.L022008, 10.1063/5.0053365, SCACCHI20251135, varma2026general}. In such systems, in bulk, the total excess free energy reduces to~\cite{likos2001effective}
\begin{equation}
F^{\mathrm{exc}}[\rho_0]
=
\frac{1}{2}\rho_0^2 AV,
\label{eq::exact_free_energy_soft}
\end{equation}
with integrated strength
\begin{equation}
A
=
4\pi
\int_0^\infty
r^2
\phi(r)\,
dr,
\label{eq::integrated_strength_soft}
\end{equation}
with $\phi(r)$ being the (unknown), soft, full interaction potential.
This can be formally obtained by setting $G(r)=1$ (which corresponds to neglecting correlations, though with smaller impact than one might expect, see~\cite{10.1063/1.4993175}) in Eq.~\eqref{eq:thermodynamic_integration} and $\phi(r)=\phi^{\rm tail}(r)$ in Eq.~\eqref{eq::potential_decomposition}. Alternatively, assuming $c_{\rm ref}^{(2)}(r) = 0$ in Eq.~\eqref{eq:bulk_free_c2_approx} provides the integrated strength
\begin{equation}
A_{c^{(2)}} = - k_{\rm B}T\, 4\pi \int r^2\, c^{(2)}(r)\,dr.
\label{eq::integrated_strength_cr_soft}
\end{equation}
Comparing Eq.~(\ref{eq::integrated_strength_soft}) with Eq.~(\ref{eq::integrated_strength_cr_soft}) recovers the random phase approximation (RPA) result, namely $\phi(r) =  - k_{\rm B}T c^{(2)}(r)$. For soft particles, the RPA has been shown to provide satisfactory results \cite{likos2001effective, camargo2011interfacial, louis2000mean, Louis20002522, varma2026general, stillinger1976phase, archer2002binary, archer2015soft}.

Equations~\eqref{eq:bulk_free_c2_approx}, \eqref{eq::integrated_strength_delta_cr} and \eqref{eq::integrated_strength_cr_soft} establish a direct link between the microscopic structure and the thermodynamics, since the DCF can be obtained from scattering experiments via the OZ relation \cite{d1991structural, ludwig2022oscillatory, wang2020quantitative}, or from simulations and optical measurements through the radial distribution function $g(r)$, as demonstrated in \cite{janai2016dipolar}.
\subsection{Relation between second virial coefficient and integrated strength}

Near the binodal of a phase-separating system, accurately determining the DCF becomes increasingly difficult, and numerical solutions of the OZ equation with standard closures often fail to converge~\cite{lutsko2007density}. To address this, we employ an alternative approach based on the second virial coefficient to estimate the integrated interaction strength, which within the present formalism serves as an initial estimate of the coexistence properties.

The second virial coefficient is defined as~\cite{hansen2013theory}
\begin{equation}
B_2 = -2\pi \int_0^\infty \left[e^{-\beta \phi(r)} - 1\right] r^2 dr.
\label{eq::second_virial_corrected}
\end{equation}
In the dilute limit, the RDF reduces to the Boltzmann factor~\cite{hansen2013theory}, given as
\begin{equation}
g_\alpha(r) \simeq e^{-\beta\phi_{\alpha}(r)}=e^{-\beta[\phi^{\mathrm{ref}}(r)+\alpha \phi^{\rm tail}(r)]}.
\label{eq::boltzmann_factor}
\end{equation}
This allows $G(r)$ to be evaluated analytically by integrating Eq.~(\ref{eq::boltzmann_factor}) with respect to $\alpha$ (as given in Eq.~(\ref{eq::g_alpha})), leading to
\begin{equation}
G(r) = e^{-\beta \phi^{\mathrm{ref}}(r)} \frac{1-e^{-\beta \phi^{\rm tail}(r)}}{\beta \phi^{\rm tail}(r)}.
\label{eq::G(r)_b2}
\end{equation}
Substituting the latter into Eq.~(\ref{eq::integrated_strength_gr}) yields
\begin{equation}
G(r)\phi^{\rm tail}(r) =
\frac{e^{-\beta \phi^{\mathrm{ref}}(r)}\left[1-e^{-\beta \phi^{\rm tail}(r)}\right]}{\beta}.
\label{eq::vfactor}
\end{equation}
By integrating as in Eq.~(\ref{eq::integrated_strength_gr}), one gets the integrated strength
\begin{equation}
\frac{4\pi}{\beta}\int_0^\infty r^2\left( e^{-\beta \phi^{\mathrm{ref}}(r)}-e^{-\beta \phi(r)}\right) dr
= 2\beta^{-1}\left(B_2 - B_{2}^{ \mathrm{ref}}\right),
\label{eq::gr_phir}
\end{equation}
where $B_{2}^{ \mathrm{ref}}$ is the virial coefficient of the reference system. Combining Eqs.~(\ref{eq::vfactor}) and (\ref{eq::gr_phir}) with Eq.~(\ref{eq::integrated_strength_gr}), we obtain
\begin{equation}
A_{B_2} = 2\beta^{-1}\left(B_2 - B_{2}^{ \mathrm{ref}}\right)
= 2\beta^{-1}\,\Delta B_2,
\label{eq::delta_b2}
\end{equation}
where $A_{B_2}$ denotes the integrated attraction strength expressed in terms of the difference in second virial coefficients between the reference and full system. As one might expect, this virial-level description becomes inaccurate at intermediate and high densities due to the increasing role of many-body correlations. 

\subsection{Colloid-polymer mixture}
\label{sec:IIG}

The formalism developed for one-component systems can be readily extended to multi-component mixtures. Here, as an example, we consider an $n$-component mixture consisting of both colloidal and polymeric species. For all pair interactions that feature a hard-core repulsion (e.g, colloid-colloid and colloid-polymer interactions) the free energy is decomposed as in Eq.~\eqref{eq:free_energy_decomposition}.

The reference system is chosen such that colloid-colloid and colloid-polymer interactions include excluded-volume effects, while polymer-polymer interactions are treated solely with the RPA. Accordingly, the excess free energy of the reference system is 
\begin{equation}
F_{\mathrm{ref}}^{\mathrm{exc}}[\{\rho_i\}] =
F_{\mathrm{hs}}[\{\rho'_i\}] + F_{\rm d}[\{\rho_i\}] + F_{\phi}[\{\rho_i\}],
\label{eq::exce_mixture}
\end{equation}
where $F_{\mathrm{hs}}$ accounts for hard-sphere correlations (colloid-colloid), $F_{\rm d}$ represents depletion effects arising from non-additivity (colloid-polymer).
Here, $\{\rho'_i\}$ denotes the subset of densities corresponding to colloidal species, while $\{\rho_i\}$ includes all components. $F_{\phi}[\{\rho_i\}]$ accounts for the contributions from the pairs interacting via purely soft potential, such as polymer-polymer, and  has the same form as in Eqs.~\eqref{eq::exact_free_energy_soft} and \eqref{eq::integrated_strength_cr_soft} for one-component systems. For pairs $i$ and $j$, the reference free energy contribution is given by
\begin{equation}
    F_{\phi}[\{ \rho_i \}] = \frac{1}{2} \sum_{i, j}  A_{ij} \rho_i \rho_j, 
\label{eq::soft_reference_part}
\end{equation}
where
\begin{equation}
A_{ij} = -k_{\mathrm{B}}T \int d\mathbf{r}\, c^{(2)}_{ij}(r).
\label{eq::delta_c_soft_mult}
\end{equation}
The total excess free energy can now be expressed as
\begin{equation}
F^{\mathrm{exc}} =
F_{\mathrm{ref}}^{\mathrm{exc}}[\{\rho_i\}] +
\frac{1}{2}\sum_{i,j} {A}_{ij}^{\rm tail}\,\rho_i \rho_j,
\label{eq::total_excess_mult}
\end{equation}
where $A_{ij}^{\rm tail}$ denotes the integrated strength of the residual tail interaction for each pair of species for which such a contribution is present. For any such pairs, wherever the decomposition is applicable (e.g., LJ potential, generalized Mie potential), the coefficient $A_{ij}^{\rm tail}$ can be expressed in terms of the difference of DCF (as given by Eq.~(\ref{eq::integrated_strength_delta_cr}) for one-component systems),
\begin{equation}
A_{ij} = -k_{\mathrm{B}}T \int d\mathbf{r}\, \Delta c^{(2)}_{ij}(r),
\label{eq::a_delta_c_mult}
\end{equation}
where $\Delta c^{(2)}_{ij}(r) = c^{(2)}_{ij}(r) - c^{(2),\mathrm{ref}}_{ij}(r)$.

A detailed evaluation of $F_{\mathrm{hs}}[\{\rho'_i\}]$, based on the Rosenfeld functional, and of $F_{\rm d}[\{\rho_i\}]$ within the free-volume theory framework, are discussed in Refs.~\citenum{ rosenfeld1989free, Lekkerkerker_1992, roth2002fundamental, brader2003statistical}.

\section{Results}

\subsection{One-component system}

To validate our approach, we first consider a one-component colloidal fluid interacting through a generalized Mie  \cite{mie1903kinetischen} potential of the form
\begin{equation}
\beta \phi^{\rm real}(r) = 4\epsilon \left[ \left(\frac{\sigma}{r}\right)^n - \left(\frac{\sigma}{r}\right)^m \right].
\label{eq::Mie}
\end{equation}
Here, $\epsilon$ controls the attraction strength, while $\sigma$ sets the size of the particles, which is set to $\sigma=1$ hereafter. We select two types of interaction. In the first case, $(n,m)=(12,6)$, which corresponds to the standard Lennard--Jones (LJ) potential, while in the second case, $(n,m)=(24,12)$, which corresponds to a steeper potential with shorter attraction. For both cases, we cut and shift the potential at $r=3.5\sigma$ and generate RDFs by means of Langevin dynamics \cite{langevin1908,brunger1984stochastic}.

\begin{figure}[t]
\includegraphics[width=\linewidth]{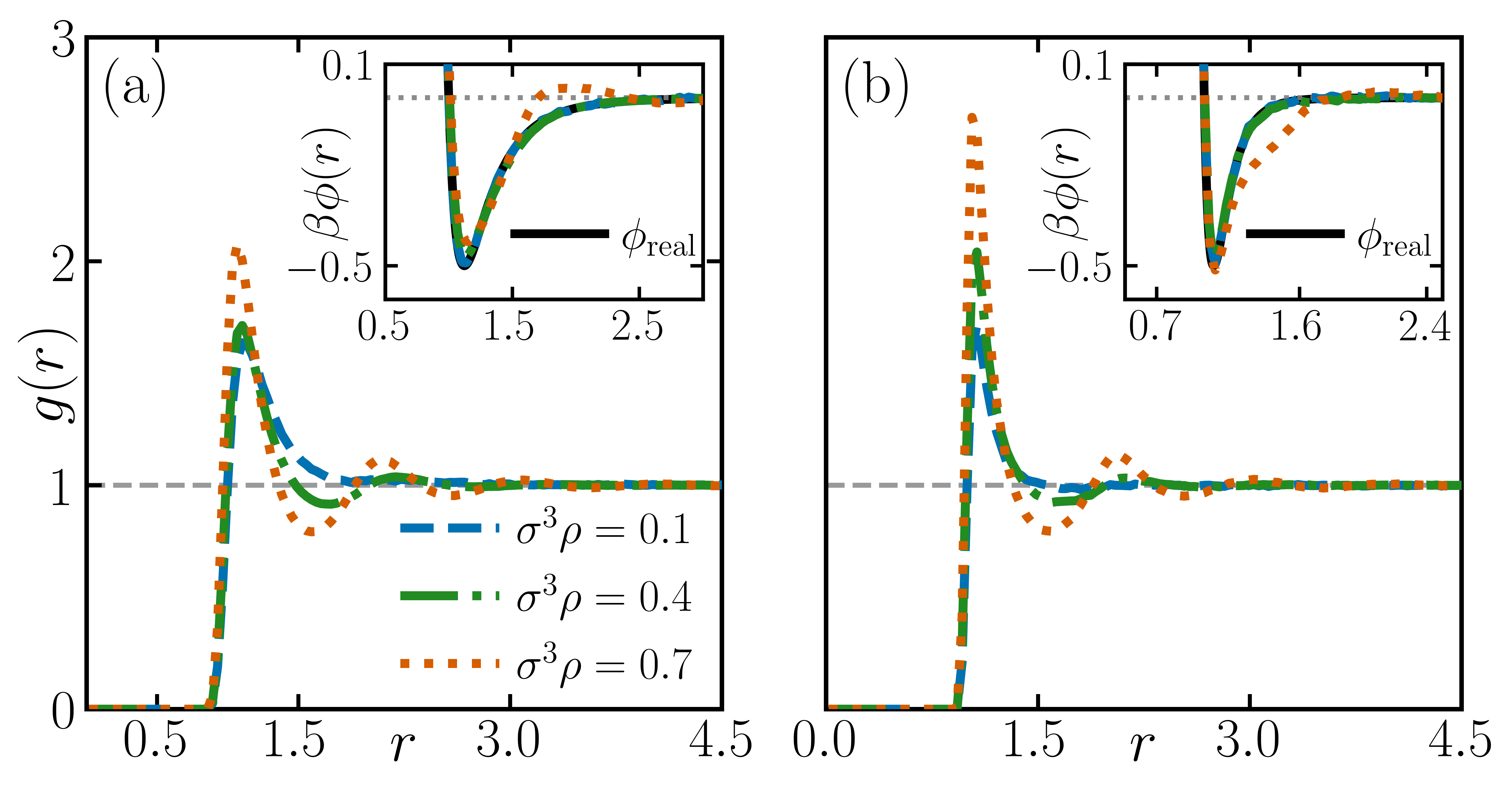}
\caption{
RDFs obtained from Langevin dynamics simulations for (a) 12--6 and (b) 24--12 Mie potential with $\beta\epsilon=0.5$ at number densities $\sigma^3\rho=0.1$, $0.4$, and $0.7$, respectively. The insets show the pair potentials used in the simulations, $\phi_{\rm real}$, together with the reconstructed potentials obtained with the IBI from the corresponding RDFs.
}
\label{fig:one_component_system}
\end{figure}

Figure~\ref{fig:one_component_system} shows the RDFs obtained from simulation and used as structural input for the 12--6 potential (panel (a)) and for the 24--12 potential (panel (b)).
These results are for $\beta\epsilon=0.5$, i.e., for temperatures above the critical temperature.
The insets compare the original pair potentials used in the simulations with the potentials reconstructed from the RDFs using the IBI procedure, where the details of the procedure are provided in the Appendix Sec.~\ref{sec:A1}.
At low and moderate densities, the reconstructed potentials are in quantitative agreement with the underlying interaction potentials. At higher densities, however, the convergence of the inversion becomes numerically slower, and the reconstructed potentials capture the real one only qualitatively. Nevertheless, this does not compromise the present approach, since the IBI potential serves only as an intermediate quantity for identifying the repulsive reference contribution and estimating the effective hard-sphere diameter entering the reference free energy. These results therefore suggest that, as a practical consideration, the structural input is preferably obtained at low-to-moderate densities, where the IBI inversion is more reliable.

\begin{figure}[t]
\includegraphics[width=\linewidth]{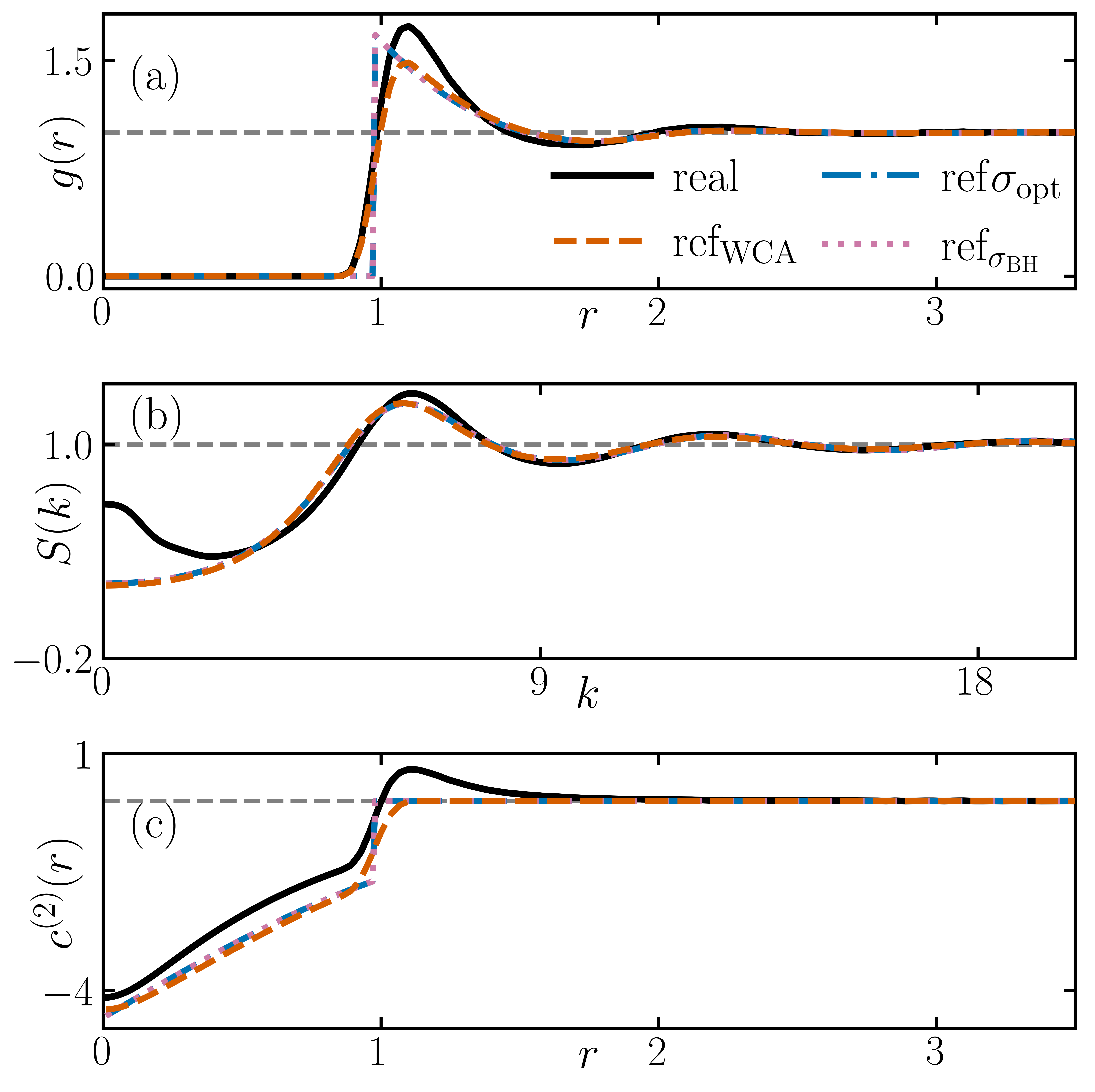}
\caption{
Structural correlation functions for a 12--6 LJ potential with $\beta\epsilon=0.5$ at number density $\sigma^3\rho=0.4$. The black solid curves correspond to the real system, with the input RDF obtained from Langevin dynamics simulations. Panel (a) shows the RDF, panel (b) the static structure factor, and panel (c) the DCF. The dashed curves correspond to the WCA reference system, obtained by retaining only the repulsive contribution of the IBI-reconstructed interaction potential. The dash-dotted curves show the effective hard-sphere reference system with diameter $\sigma_{\rm opt}$ [see Eq.~\eqref{eq:diameter_fitting}], while the short-dashed curves correspond to the hard-sphere reference system with the diameter determined from the BH prescription.
}
\label{fig:structural_correlation_one_component_system}
\end{figure}

We next examine how the structural information is mapped onto the reference system to determine the effective particle diameter. In experiments, the static structure factor, $S(k)$, is often directly accessible, while the corresponding radial distribution function, $g(r)$, can be obtained through the standard Fourier-transform relation between $S(k)$ and $g(r)$ \cite{hansen2013theory}. In simulations, the radial distribution function is readily computed from particle configurations, from which the static structure factor can subsequently be obtained.
To get the reference system, first, we decompose the inverted potential using the WCA scheme, allowing us to isolate the repulsive reference contribution, which is then mapped onto an effective hard-sphere system through the diameter-fitting procedures discussed in Sec.~\ref{sec:IIA}.
Figure~\ref{fig:structural_correlation_one_component_system} compares the structure factor, RDF, and DCF of the 12--6 system with those of the corresponding reference systems. These include the WCA reference system, constructed from the repulsive part of the inverted IBI potential, and two hard-sphere approximations based on either the BH diameter [Eq.~\eqref{eq::sigma_bh}] or the optimized diameter, $\sigma_{\rm opt}$, obtained by fitting the RDF using Eq.~\eqref{eq:diameter_fitting}. The corresponding reference DCFs are calculated by solving the OZ equation with the PY closure. The main differences between the real and reference systems arise from the contribution of the attractive tail, which is absent from the reference descriptions. The close agreement among the WCA, BH, and optimized hard-sphere references indicates that the excluded-volume contribution is largely insensitive to the particular reference construction, supporting the separation of the excess free energy into reference and tail contributions.

\begin{figure}[t]
\includegraphics[width=\linewidth]{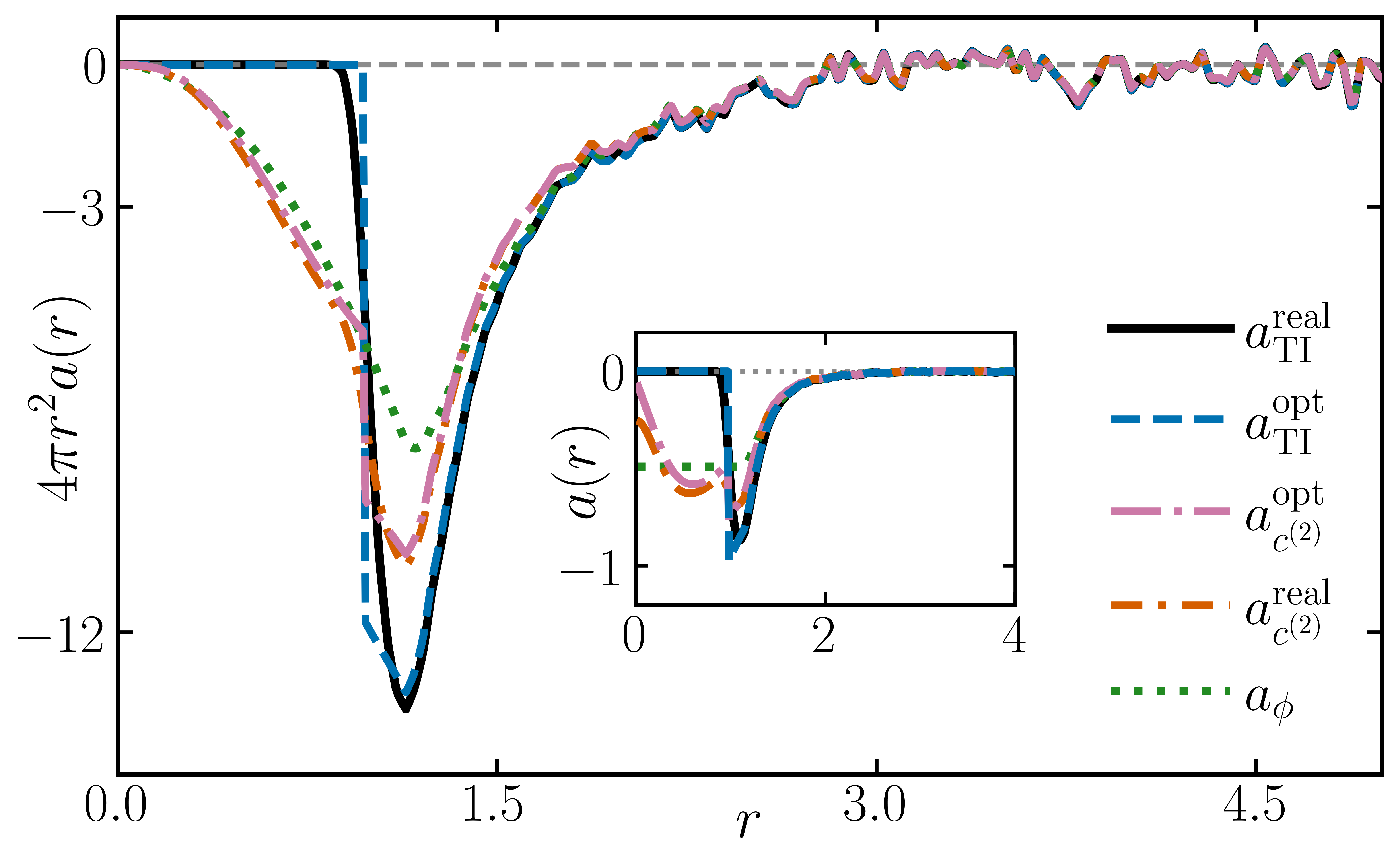}
\caption{
Kernel $a(r)$, as defined in Eqs.~\eqref{eq::integrated_strength_gr} and \eqref{eq::integrated_strength_delta_cr}, for a 12--6 LJ potential with $\beta\epsilon=0.5$ at number density $\sigma^3\rho=0.4$, obtained using the different routes. The solid lines show the TI result using the WCA reference, while the long-dashed lines show the corresponding result using the optimized hard-sphere reference. The long dash-dotted lines show the $\Delta c^{(2)}$ result using the optimized hard-sphere reference, and the short dash-dotted lines correspond to the WCA-based repulsive reference. Finally, the dotted lines, $a_{\phi}$, represent the attractive tail obtained by splitting the bare IBI-reconstructed potential [cf.~Eq.~\eqref{eq::integrated_strength_soft}] for the simulated RDF.
}
\label{fig:kernel_integrated_strength_one_component}
\end{figure}
$F^{\rm exc}_{\rm ref}$ can be evaluated using, e.g., the White Bear version of FMT \cite{Roth_2010, roth2002fundamental, hansen2006new}, while $F^{\rm exc}_{\rm tail}$ depends on the integrated interaction strength $A$, which is obtained by integrating the kernel $a(r)$; see Eqs.~\eqref{eq::integrated_strength_gr} and \eqref{eq::integrated_strength_delta_cr}. Figure~\ref{fig:kernel_integrated_strength_one_component} compares the kernels $4\pi r^2 a(r)$ obtained from the different routes discussed in Sec.~\ref{sec:II}; the corresponding $a(r)$ functions are shown in the insets. The DCF-based kernels, obtained using either the WCA reference, $a_{c^{(2)}}^{\rm real}$, or the optimized hard-sphere reference, $a_{c^{(2)}}^{\rm opt}$, remain close to those obtained from TI using the corresponding reference systems, $a_{\rm TI}^{\rm real}$ and $a_{\rm TI}^{\rm opt}$ [see Eq.~\eqref{eq:bulk_free_c2_approx}]. This agreement indicates that the attractive contribution extracted from the structural correlations is relatively insensitive to the particular representation of the repulsive reference system. In contrast, the kernel obtained directly from the attractive part of the WCA-decomposed potential shows substantially larger deviations. This difference highlights that the attractive contribution entering the free energy is not, in general, equivalent to the bare attractive tail of the interaction potential, but depends on the structural response of the fluid to that interaction. The agreement between the DCF and TI routes therefore supports the use of the correlation-based kernel as the relevant quantity for the free-energy calculation.

\begin{figure}[t]
\includegraphics[width=\linewidth]{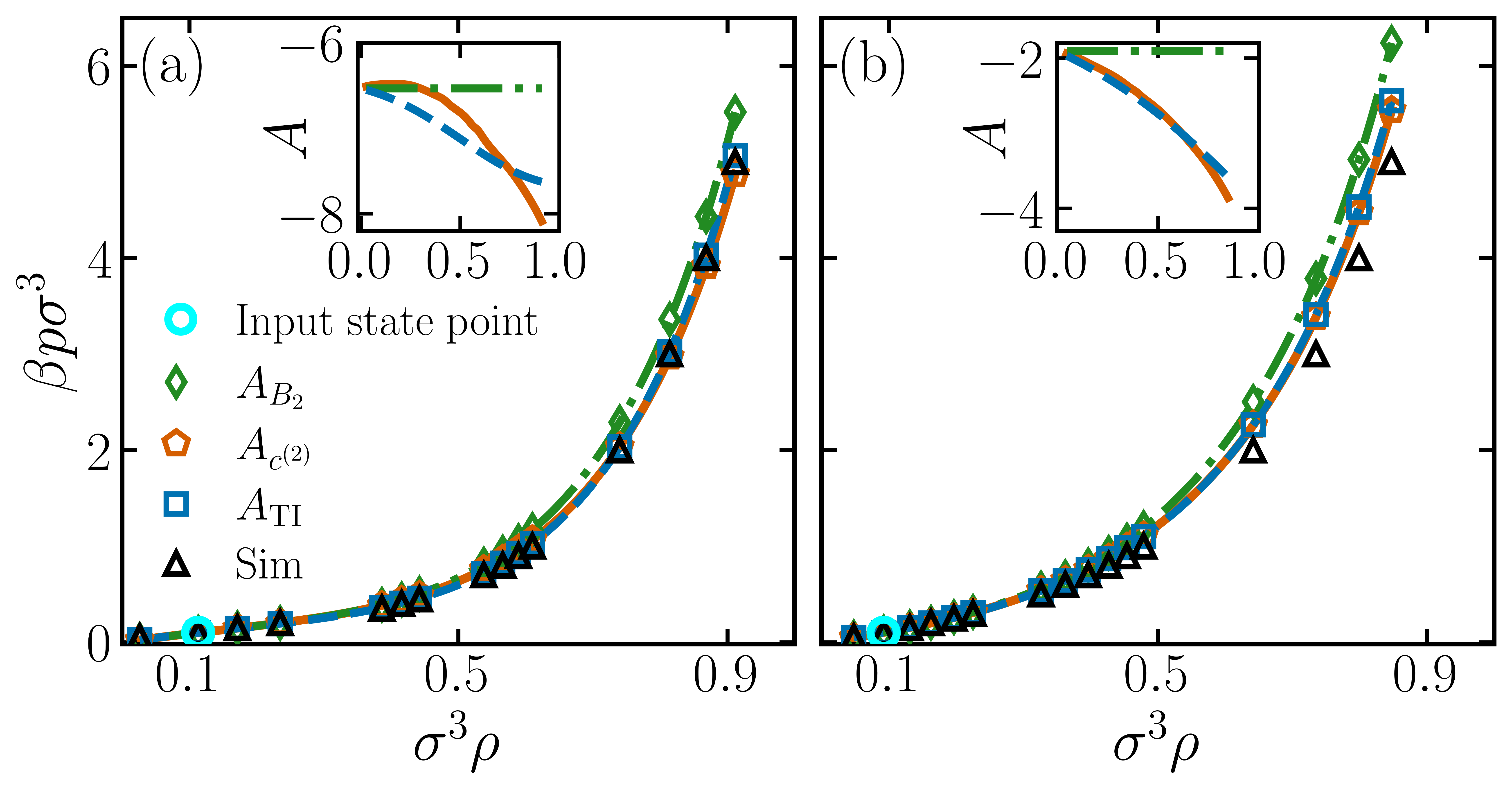}
\caption{
Equation of state for a one-component fluid interacting through (a) the Lennard–Jones 12–6 potential and (b) the generalized Mie 24–12 potential, both with $\beta\epsilon = 0.5$. Triangles denote simulation results, diamonds the virial route, squares the TI route, and circles the DCF route. The inset shows the corresponding density dependent integrated strength.
}
\label{fig::colloid}
\end{figure}

To assess the accuracy of the free energies obtained from the different routes, Figs.~\ref{fig::colloid}(a) and (b) compare the corresponding equations of state with direct simulation results for the 12--6 and 24--12 Mie potentials, respectively. The only structural input is obtained at the single state point marked by the open circle. From this input, the RDF is used to construct the reference description, estimate the effective particle diameter, and determine the integrated strength, which is then used to predict the pressure over a range of densities.

For both systems, the TI [Eq.~\eqref{eq::exact_free_energy_final}] and DCF [Eq.~\eqref{eq::integrated_strength_delta_cr}] routes show substantially better agreement with simulation than the virial route. This is expected because the virial route is based on a dilute-limit description and therefore does not fully capture the density dependence of the effective attractive contribution. This limitation is evident in the inset of Fig.~\ref{fig::colloid}, where the integrated strength obtained from the virial route is independent of density. In contrast, the TI and DCF routes incorporate structural information beyond the dilute limit, resulting in more accurate equations of state at moderate and high densities.

\begin{figure}[t]
\includegraphics[width=\linewidth]{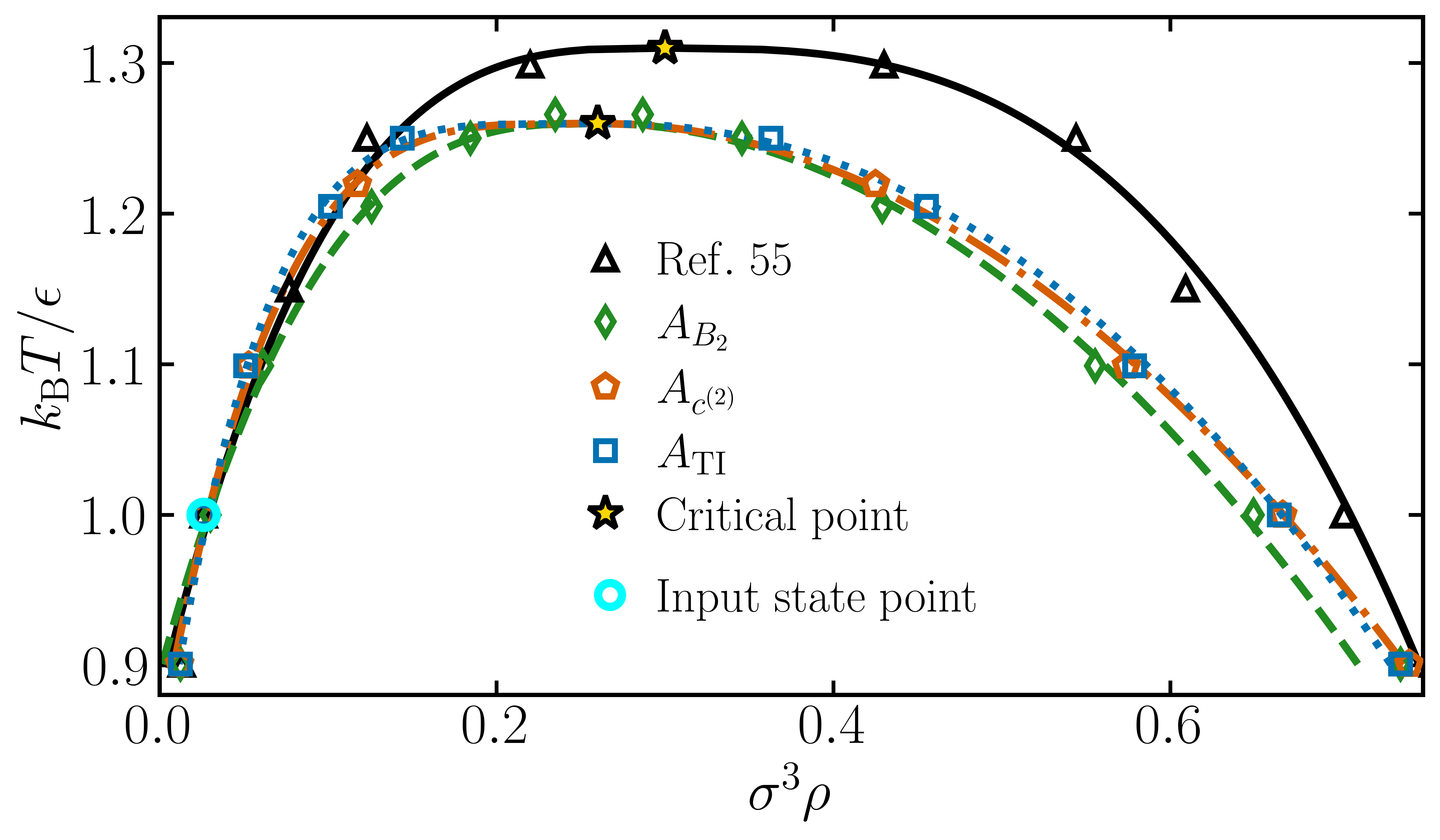}
\caption{
Binodal curves for the 12--6 system obtained using the different routes, compared with simulation data from Ref.~\citenum{Panagiotopoulos01071987}. Stars denote the critical points. The open circle marks the single state point, $(\sigma^3\rho,k_{\rm B}T\epsilon^{-1})=(0.05,1)$, used as structural input.
}
\label{fig::binodal}
\end{figure}

We also test the consistency of our approach by comparing the binodal obtained for the 12–6 potential using the TI, DCF, and virial routes with the simulation results reported in Ref.~\citenum{Panagiotopoulos01071987}. The comparison is shown in Fig.~\ref{fig::binodal}. Overall, we find good agreement, particularly for the gas branch of the binodal. However, the simulations reported in Ref.~\citenum{Panagiotopoulos01071987} include a tail correction, whereas our calculations employ a truncated-and-shifted potential. The critical point is slightly underestimated, as is much of the liquid branch. Some part of this discrepancy may therefore be attributed to differences in the treatment of the interaction potential. Nevertheless, all three routes reproduce the overall shape of the coexistence curve, with the TI route showing the best agreement, followed by the DCF route and, finally, the virial route.

It is also important to note that the critical points obtained from the three routes are closely related because all calculations are initialized from the coexistence curve obtained using the virial route. The TI and DCF routes subsequently refine the coexistence densities, resulting in some variation in the predicted tie lines. However, as the coexistence curve approaches its termination, the initial virial estimate determines the limiting point of the calculation. Consequently, the tie lines terminate at nearly the same location for all three routes, giving closely overlapping estimates of the critical point.

Taken together, the results so far demonstrate that accurate thermodynamic properties and phase behaviour can be recovered directly from structural correlations without requiring explicit interaction potentials. Moreover, near the critical point and within the coexistence region, where homogeneous structural data become increasingly difficult to obtain, the smooth density dependence of the integrated strength $A$ permits reliable interpolation or extrapolation into these regions, further extending the applicability of the proposed framework.

\subsection{Two-component colloid-polymer mixture}

\begin{table*}[t]
\centering
\begin{tabular}{c|c|c|c}
\hline\hline
Pair & Potential & Case I & Case II \\
\hline
cc
& Mie [Eq.~(\ref{eq::Mie})]
& $\beta\epsilon_{\rm cc}=0.5,\;
   \sigma_{\rm cc}=\sigma,\;
   (n,m)=(12,6),\;
   r_{\rm c}=3.5\sigma$
& $\beta\epsilon_{\rm cc}=0.5,\;
   \sigma_{\rm cc}=\sigma,\;
   (n,m)=(24,12),\;
   r_{\rm c}=3.5\sigma$\\[2mm]

cp
& Mie [Eq.~(\ref{eq::Mie})]
& $\beta\epsilon_{\rm cp}=1.0,\;
   \sigma_{\rm cp}=\sigma,\;
   (n,m)=(12,6),\;
   r_{\rm c}=2^{1/6}\sigma_{\rm cp}$
& $\beta\epsilon_{\rm cp}=\sigma,\;
   \sigma_{\rm cp}=\sigma,\;
   (n,m)=(12,6),\;
   r_{\rm c}=2^{1/6}\sigma_{\rm cp}$
\\[2mm]

pp
& Gaussian [Eq.~(\ref{eq::gaussian})]
& $\beta\epsilon_{\rm pp}=2.0,\; R_{\rm pp}=\sigma$
& $\beta\epsilon_{\rm pp}=2.0,\; R_{\rm pp}=\sigma$
\\
\hline\hline
\end{tabular}

\caption{
Pair interactions used for the two-component colloid--polymer mixtures.
The variable $r_{\rm c}$ corresponds to the cutoff distance after which the interaction is set to zero.
}
\label{tab:binary_interactions}
\end{table*}

We now extend the approach to a binary colloid--polymer mixture. This provides a more stringent test of the framework, as the thermodynamics depends not only on the correlations within each component but also on the cross correlations between colloids and polymers. We therefore ask
whether structural information obtained at a single thermodynamic state point is sufficient to determine the integrated interaction strengths, $A_{ij}$, and thereby predict the equation of state and phase behavior of the mixture.

For the purpose, we consider two different systems. The pair interactions parameters considered for the binary mixtures are summarized in Table~\ref{tab:binary_interactions}. In both cases, the polymer--polymer interaction is described by a Gaussian potential \cite{PhysRevE.64.041501,Bolhuis20014296},

\begin{equation}
\beta \phi_{\rm pp}(r)=
\epsilon_{\rm pp}
\exp\left(-\frac{r^2}{R_{\rm pp}^2}\right),
\label{eq::gaussian}
\end{equation}

while the colloid--polymer interaction is represented by a purely repulsive Weeks--Chandler--Andersen (WCA) potential \cite{weeks1971role}. The two cases differ only in the colloid--colloid interaction. Case I employs a 12--6 generalized Mie potential, whereas Case II employs a 24--12 generalized Mie potential. In both cases, the colloid--colloid interaction contains an attractive contribution.
\begin{figure*}[!t]
\includegraphics[width=\linewidth]{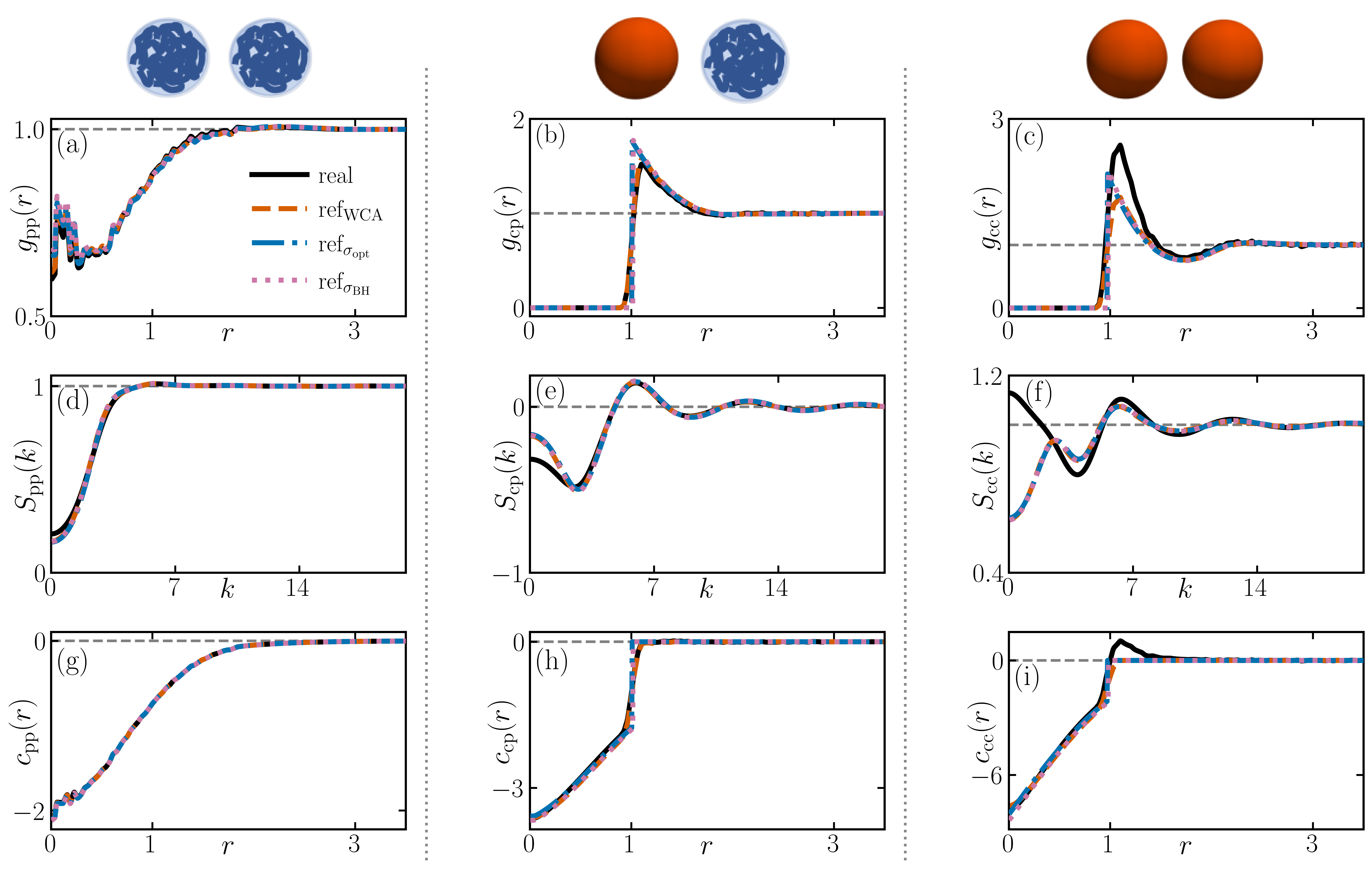}
\caption{
Structural and correlation properties of the two-component colloid--polymer system for Case I of Table~\ref{tab:binary_interactions}. The left, middle, and right columns correspond to polymer--polymer, colloid--polymer, and colloid-colloid correlations, respectively, while the rows show the RDFs, static structure factors, and DCFs. Solid curves denote the real system obtained from simulation and used as structural input. Dashed curves show the WCA reference system, dash-dotted curves the optimized hard-sphere reference with diameter $\sigma_{\rm opt}$, and dotted curves the hard-sphere reference with BH diameter $\sigma_{\rm BH}$. The densities are fixed at $\sigma^3\rho_{\rm p}=0.6$ and $\sigma^3\rho_{\rm c}=0.09$.
}
\label{fig:structural_correlation_two_component_system}
\end{figure*}
For both the cases, the input radial distribution functions are obtained from Langevin dynamics simulations at polymer and colloid number densities $\sigma^3\rho_{\rm p}=0.6$ and $\sigma^3\rho_{\rm c}=0.09$, respectively. As in the one-component case, the IBI procedure is used only as an intermediate step to construct a state-point-specific decomposition of the interactions and to determine the corresponding excluded-volume reference contributions and therefore the size of the particles.

For the present system, the nature of each pair interaction is inferred from the input radial distribution functions. The polymer--polymer interaction is purely soft, whereas the colloid--polymer and colloid--colloid interactions contain excluded-volume contributions. We therefore construct reference descriptions for the latter two interactions using the WCA decomposition of the repulsive core and subsequently map them onto effective hard-sphere interactions. For the hard-sphere mapping, the diameter is determined either from the optimized RDF fit, $\sigma_{\rm opt}$, or from the BH prescription, $\sigma_{\rm BH}$. The resulting reference descriptions are compared with the structural correlations of the full system in Fig.~\ref{fig:structural_correlation_two_component_system}.

More generally, when the nature of the underlying interactions is not known a priori, the structural correlations themselves can provide guidance in selecting the appropriate reference description and so the length scale. A finite RDF at short separations is indicative of a soft, inter-penetrable interaction, whereas a region in which $g_{ij}(r)=0$ signals an excluded-volume contribution that can be mapped onto a hard-sphere-like reference. Thus, the distinction between soft and excluded-volume interactions can, in principle, be inferred directly from the structural input rather than from prior knowledge of the microscopic interaction potentials.

For the polymer--polymer pair [Fig.~\ref{fig:structural_correlation_two_component_system}(a)], the interaction is purely soft and is retained entirely in the reference description; consequently, the reference and real-system RDFs coincide. The colloid--polymer interaction [Fig.~\ref{fig:structural_correlation_two_component_system}(b)] is purely repulsive, so only minor differences arise between the real and reference systems. In contrast, removing the attractive colloid--colloid contribution produces a more noticeable change in $g_{\rm cc}(r)$ [Fig.~\ref{fig:structural_correlation_two_component_system}(c)]. Nevertheless, the principal peak positions remain nearly unchanged, indicating that the attraction primarily modifies the strength rather than the characteristic length scale of the correlations. The weak response of the polymer--polymer correlations to the attractive colloid--colloid interaction further indicates that the polymer distribution is only weakly affected by this contribution. This is consistent with the reservoir-like role of the polymer component underlying the free-volume description~\cite{Lekkerkerker_1992}. The effect of the reference decomposition becomes more pronounced in reciprocal space. The corresponding static structure factors are shown in Fig.~\ref{fig:structural_correlation_two_component_system}(d)--(f). While $S_{\rm pp}(k)$ [Fig.~\ref{fig:structural_correlation_two_component_system}(d)] remains relatively insensitive to the reference description, more pronounced differences appear in $S_{\rm cp}(k)$ and $S_{\rm cc}(k)$ [Fig.~\ref{fig:structural_correlation_two_component_system}(e) and (f)], particularly at small $k$. This enhanced sensitivity reflects the role of the colloid--colloid attraction in collective density fluctuations and, consequently, in the phase behaviour of the mixture (discussed below). A similar trend is observed for the direct correlation functions in Fig.~\ref{fig:structural_correlation_two_component_system}(g)--(i). The polymer--polymer DCF, $c_{\rm pp}^{(2)}(r)$, is unchanged because the entirety of the soft interaction is included in the reference description, while only small differences occur for $c_{\rm cp}^{(2)}(r)$ in the absence cross-attraction. In contrast, $c_{\rm cc}^{(2)}(r)$ shows a clear deviation between the real and reference systems, isolating the contribution of the attractive colloid--colloid interaction that enters the integrated-strength correction discussed below.

Once the effective hard-sphere diameters, $\sigma_{\rm opt}^{\rm cc}$ and $\sigma_{\rm opt}^{\rm cp}$, have been determined, the excess free energy of the reference mixture can be constructed using Eq.~\eqref{eq::exce_mixture}. The hard-sphere contribution, given by the first term of Eq.~\eqref{eq::exce_mixture}, is evaluated using the White Bear Mark I functional for hard-sphere mixtures~\cite{Roth_2010}, while the colloid--polymer depletion contribution [second term of Eq.~\eqref{eq::exce_mixture}] is described using multicomponent free-volume theory~\cite{Lekkerkerker_1992}. The effective polymer size entering this contribution is determined from the additive relation between the characteristic length scales of the colloid--colloid and colloid--polymer interactions, which is expressed by the relation, $\sigma_{\rm pp} = 2 \sigma_{\rm cp} -  \sigma_{\rm cc}$.

The remaining polymer--polymer contribution is treated directly using the soft-particle formulation introduced in Sec.~\ref{sec:IIG}, with its thermodynamic contribution evaluated from the DCF or virial route [Eq.~\eqref{eq::integrated_strength_cr_soft}]. Thus, the reference free energy combines the hard-sphere colloidal contribution, the colloid--polymer depletion contribution, and the explicitly treated soft polymer--polymer interaction.

Having determined the reference free energy, the total excess free energy of the interacting mixture follows from Eq.~\eqref{eq::total_excess_mult}. The remaining quantities are the integrated tail strengths, $A_{ij}^{\rm tail}$, for pair interactions containing an excluded-volume reference. For such pairs, $A_{ij}^{\rm tail}$ is obtained from the difference between the DCFs of the full and reference systems according to Eq.~\eqref{eq::a_delta_c_mult}, or alternatively from the virial or TI routes following the procedures introduced for the one-component system.

\begin{figure}[h]
\includegraphics[width=\linewidth]{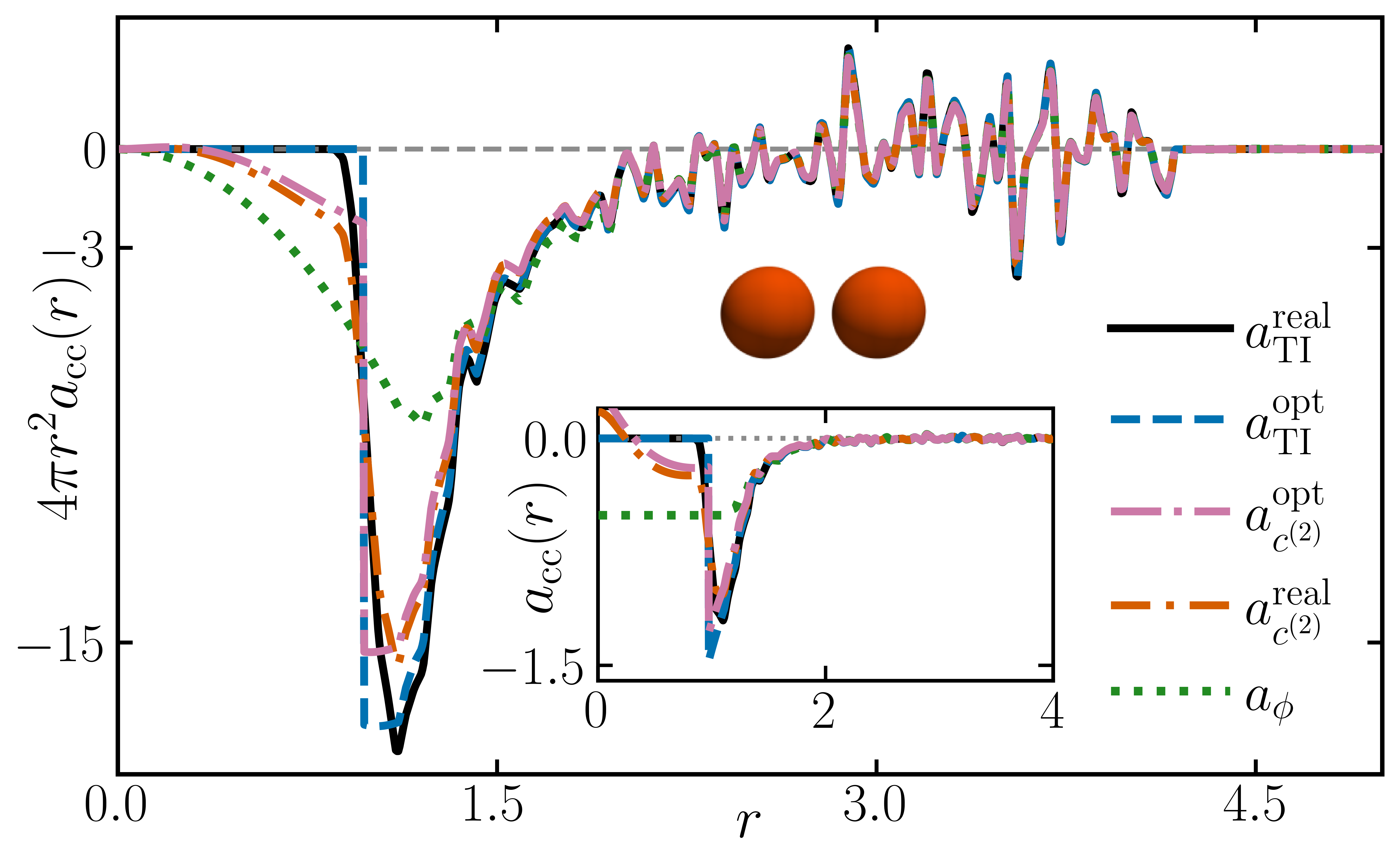}
\caption{
Kernel $a_{\rm cc}(r)$, the integrand in Eq.~\eqref{eq::a_delta_c_mult}, for the colloid--colloid interaction of Case I in Table~\ref{tab:binary_interactions}. The solid lines show the TI result using the WCA reference, while the long-dashed lines show the corresponding result using the optimized hard-sphere reference. The long dash-dotted lines show the $\Delta c^{(2)}$ result using the optimized hard-sphere reference, and the short dash-dotted lines correspond to the WCA-based repulsive reference. Finally, the dotted lines, $a_{\phi}$, represent the attractive tail obtained by splitting the bare IBI-reconstructed potential [cf.~Eq.~\eqref{eq::integrated_strength_soft}] for the simulated RDF.
}
\label{fig:kernel_integrated_strength_two_component}
\end{figure}

For the present mixture, the only nonzero tail contribution associated with an excluded-volume reference arises from the colloid--colloid interaction. Although the colloid--polymer pair contains an excluded-volume contribution, it has no additional tail, and hence $A_{\rm cp}^{\rm tail}=0$. Figure~\ref{fig:kernel_integrated_strength_two_component} shows the corresponding colloid--colloid kernel, $a_{\rm cc}(r)$, in the inset, while the main panel displays the weighted kernel $4\pi r^2a_{\rm cc}(r)$ whose integral gives $A_{\rm cc}^{\rm tail}$. The TI and DCF routes give closely agreeing results, and the kernels obtained using the WCA and optimized hard-sphere references are likewise very similar. This indicates that the extracted colloid--colloid tail contribution is only weakly sensitive to the particular representation of the repulsive core. In contrast, the hard-core contribution to the reference free energy remains sensitive to the effective particle diameter~\cite{baxter1964direct}.

\begin{figure}[t]
\includegraphics[width=\linewidth]{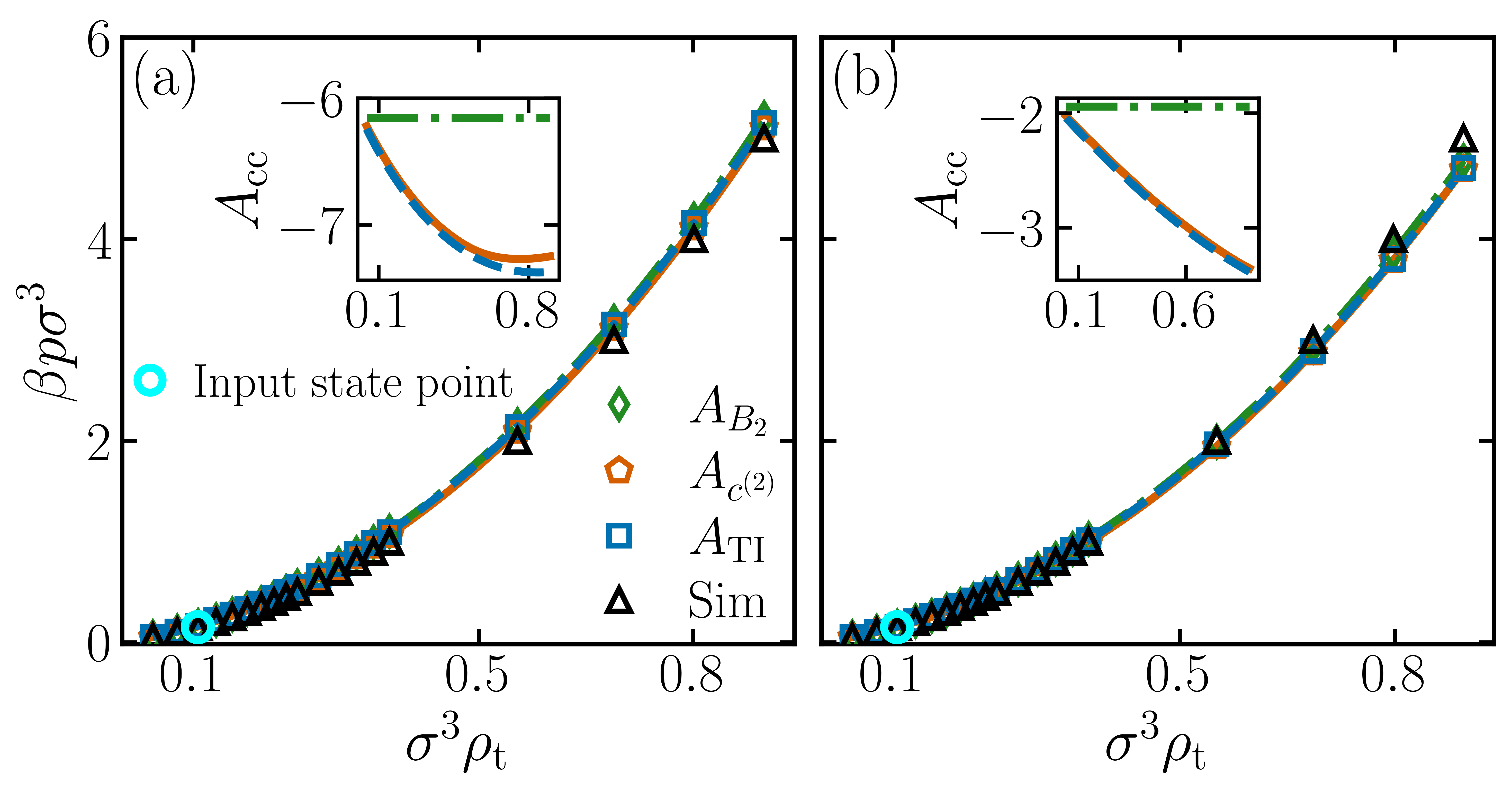}
\caption{
Equation of state for the colloid--polymer mixture, shown as pressure versus total density $\rho_{\rm t}$ at fixed colloid fraction
$x=0.16$. The pressure is computed from the integrated strengths obtained using the different routes indicated in the legend.
Panels (a) and (b) correspond to Cases I and II of Table~\ref{tab:binary_interactions}, respectively. The open circular marker indicates the single simulation state point used as structural correlation input.
}
\label{fig::colloid_polymer}
\end{figure}

We next use the structural input to construct the free energy and predict the equation of state for the two mixtures defined in Table~\ref{tab:binary_interactions} (Case I in panel (a) and Case II in panel (b)). In both cases, the pressure is evaluated as a function of total density, $\rho_{\rm t}$, at fixed colloid fraction $x=0.16$.

The structural input, obtained from Langevin dynamics simulations and marked by the open symbol, is used to determine the matrix of integrated interaction strengths, $A_{ij}$. The resulting free-energy description is then used to predict the pressure over the full density range, without requiring structural data at each density. For both mixtures, the TI and DCF routes show very good agreement with the simulation data. The virial route also performs better than for the one-component attractive fluid, which is consistent with the smaller colloid fraction and hence the weaker contribution of the attractive colloidal interactions to the total pressure. Nevertheless, the TI and DCF routes remain more reliable. The weak and moderate density dependence of the integrated strength shown in the inset further explains the robustness of the structural-correlation-based predictions for these mixtures.
\begin{figure}[h!]
\includegraphics[width=\linewidth]{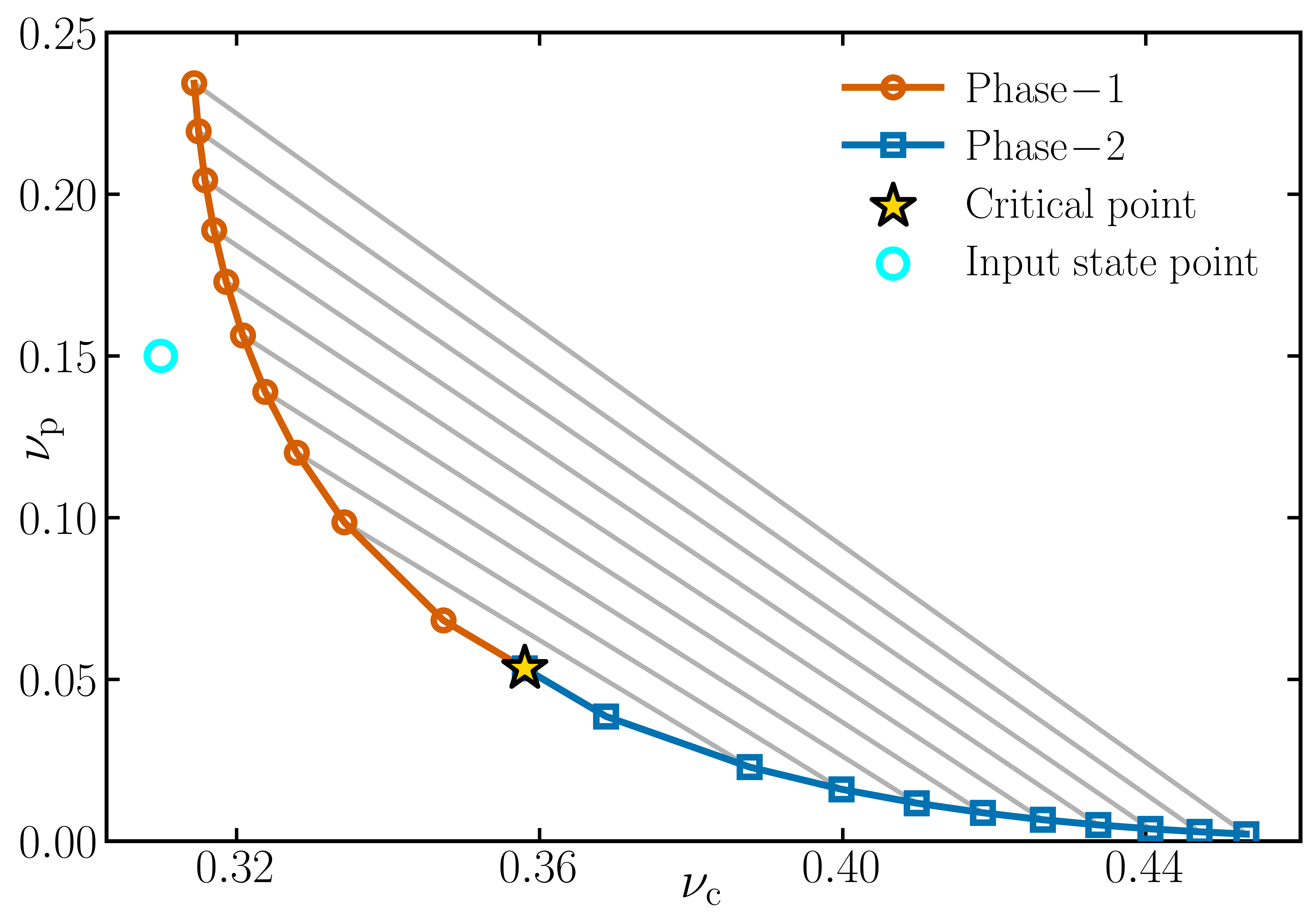}
\caption{
Binodal curve for the binary colloid--polymer mixture corresponding to Case II of Table~\ref{tab:binary_interactions}, predicted from structural correlations obtained at the input state point indicated by a large open circle. The integrated strengths entering the free-energy functional are computed using the DCF route. The two branches denote the coexisting polymer-rich/colloid-poor and colloid-rich/polymer-poor fluid phases.
}
\label{fig::binodal_binary}
\end{figure}
Finally, we apply our approach to predict the fluid--fluid coexistence of the binary colloid--polymer mixture corresponding to Case II in Table~\ref{tab:binary_interactions}. The binodal is obtained by solving the coexistence conditions~\cite{hansen2013theory}. Structural correlations at the single input state point are used to evaluate the $\Delta c^{(2)}$-based integrated strengths, which are then used to construct the binary free energy and determine the coexistence curve. The resulting binodal is shown in Fig.~\ref{fig::binodal_binary} in terms of the occupied volume fractions of the two components, defined as $\nu_i = \rho_i v_i,$ where $\rho_i$ is the number density of species $i$ and $v_i$ is the characteristic volume associated with that species. For the colloids, $v_{\rm c}$ is defined from the effective hard-sphere diameter, while for the polymers, which are represented by penetrable Gaussian particles, $v_{\rm p}$ is defined from their characteristic size $R_{\rm pp}$. The two branches correspond to coexisting phases with different polymer and colloid concentrations, demonstrating that the structural correlations contain sufficient information to predict demixing even when the polymers are nonideal and interact through a soft repulsive potential. This is particularly relevant because classical descriptions of colloid--polymer demixing often rely on ideal-polymer assumptions, which become inappropriate when polymer--polymer interactions are present. Here, this limitation is avoided by constructing the free energy directly from the structural correlations.

The calculated binodal should be interpreted as a fluid--fluid coexistence prediction within the homogeneous-fluid regime. At sufficiently high colloid densities, a fluid--solid transition may become relevant, which is beyond the scope of the present fluid-only formulation. Nevertheless, the result demonstrates that structural information obtained at a single state point can be used to construct the free energy and predict phase coexistence in an interacting colloid--polymer mixture.

\section{Conclusion}

In this work, we have developed a structural-correlation-based framework for predicting the thermodynamics and phase behaviour of colloidal and colloid--polymer mixtures. The central idea is to construct the excess free energy directly from structural correlations, in the form of radial distribution functions, static structure factors, and direct correlation functions, rather than from explicitly known interaction potentials. This provides a route for studying soft-matter systems in which effective interactions may be unknown, state-point dependent, or affected by many-body effects.

The formalism separates the contribution of short-range excluded-volume interactions from that of the remaining interaction. Interactions containing a repulsive core are mapped onto an effective hard-sphere reference system, while the remaining contribution is described through an integrated interaction strength, $A$. This quantity can be evaluated using thermodynamic integration, the direct correlation function route, or the virial route. Inverse Boltzmann iteration is used only as an intermediate step to construct the state-point-specific reference decomposition and estimate its excluded-volume contribution; the thermodynamic predictions themselves follow from the resulting free-energy construction.

We first tested the framework for two one-component colloidal systems. Using structural information from a single thermodynamic state point, we predicted the equation of state over a broad range of densities, obtaining good agreement with simulation data. The thermodynamic integration and direct correlation function routes performed substantially better than the virial route, particularly at moderate and high densities, where the dilute-limit approximation underlying the latter becomes less accurate. We further predicted the fluid--fluid binodal of one of these systems, showing that the overall coexistence behaviour can be recovered from structural information at a single state point.

We then extended the framework to binary colloid--polymer mixtures, where both self- and cross-correlations contribute to the thermodynamics. Structural information obtained at a single state point was used to construct the integrated-strength matrix, $A_{ij}$, and predict the equation of state at fixed composition. The resulting predictions show good agreement with simulation data, demonstrating that the framework can describe mixtures containing both soft polymeric and excluded-volume colloidal components without requiring explicit knowledge of their interaction potentials.

Finally, we applied the framework to a different colloid--polymer mixture to predict its fluid--fluid coexistence. No comparison with simulation was made in this case; rather, the calculation demonstrates that the framework can generate a phase diagram from structural information at a single homogeneous state point. Since the free-energy functional is constructed directly from structural correlations, the approach does not require an explicit interaction potential or a system-specific thermodynamic model. This provides a potential route toward determining the phase behaviour of complex fluids from experimentally accessible structural information, even when their microscopic interactions are not known.

Overall, the present framework provides a general route for constructing free-energy functionals directly from structural correlations. Its key feature is that thermodynamic and phase behaviour can be inferred from structural information obtained at a single thermodynamic state point, without requiring explicit knowledge of the underlying interaction potentials or extensive thermodynamic integrations. This is particularly relevant for experimental systems, where the effective interactions between components are often difficult to determine accurately and may depend on the state of the system. The present formulation is developed for isotropic interactions and therefore does not explicitly account for orientational degrees of freedom. Nevertheless, it may remain applicable to some anisotropic systems when the structural correlations are sufficiently well described by their isotropic components, for example in regimes where no strong orientational ordering occurs. Similarly, strongly correlated systems, such as highly charged colloidal or ionic suspensions, may require a more sophisticated treatment if pair correlations alone are insufficient to describe their thermodynamics. Whether and to what extent the present framework remains accurate in such systems is an open question that can be addressed by further testing against simulation and experiment. Extensions to incorporate anisotropy and higher-order correlation effects therefore provide natural directions for future work.

\section*{Acknowledgments}
This work was supported by the Jane and Aatos Erkko Foundation under the grant No. 230052 (AS). Computational resources by CSC IT Center for Finland are also gratefully acknowledged. The work was conducted within the \#SUSMAT profiling measure, at the University of Turku, Finland.

\appendix 
\section*{APPENDIX} 
\setcounter{section}{0}
\renewcommand{\thesubsection}{A.\arabic{subsection}}
\setcounter{equation}{0}
\renewcommand{\theequation}{A.\arabic{equation}}
\setcounter{figure}{0}
\renewcommand{\thefigure}{A.\arabic{figure}}

\subsection{Iterative Boltzmann inversion}
\label{sec:A1}

When the interaction potential and particle size are not known a priori, integral-equation theory provides a route for inferring an effective pair potential from the radial distribution function. In the low-density limit, the potential is related to the isotropic radial distribution function through
\begin{equation}
\beta\phi(r)=-\ln g(r).
\end{equation}
At finite density, however, this relation neglects many-body correlations. We therefore use IBI,
\begin{equation}
\beta \phi^{(n)}(r)
=
\beta \phi^{(n-1)}(r)
+
\gamma
\ln\left[
\frac{g^{(n-1)}(r)}
{g^{\mathrm{target}}(r)}
\right],
\label{eq::ibi_update}
\end{equation}
where $n$ is the iteration number, $\gamma$ is a mixing parameter controlling convergence, and $g^{\mathrm{target}}(r)$ is the target radial distribution function, obtained from either scattering experiments or molecular simulations.

At each iteration, the radial distribution function corresponding to the trial potential is calculated by solving the OZ equation [Eq.~\eqref{eq::oz_real_space_1c}] with an appropriate closure. We use the PY closure for interactions containing a hard core and the hypernetted-chain closure for soft interactions. The resulting total correlation function, $h(r)$, gives $g(r)=1+h(r)$, which is then used to update the potential according to Eq.~\eqref{eq::ibi_update}. The procedure is repeated until the calculated radial distribution function converges to the target.

Here, the IBI procedure is applied to the two-component colloid--polymer system, using the radial distribution functions obtained from molecular dynamics simulations as the target structural input. For the multicomponent system, the self-interaction potentials are first reconstructed from the corresponding one-component radial distribution functions, treating the colloid and polymer components independently. These potentials are then kept fixed while the colloid--polymer potential is reconstructed from the cross radial distribution function. Thus, as the most reliable strategy, a complete reconstruction requires the three pair correlation functions, $g_{\rm cc}(r)$, $g_{\rm pp}(r)$, and $g_{\rm cp}(r)$, evaluated at the same thermodynamic state point and obtained from three separate experiments: one on the pure polymer system, one on the pure colloidal system, and one on the colloid--polymer mixture. Note, that this strategy assumes that the self-interactions of the individual components remain unchanged upon varying the mixture composition.

\begin{figure}[t]
\includegraphics[width=\linewidth]{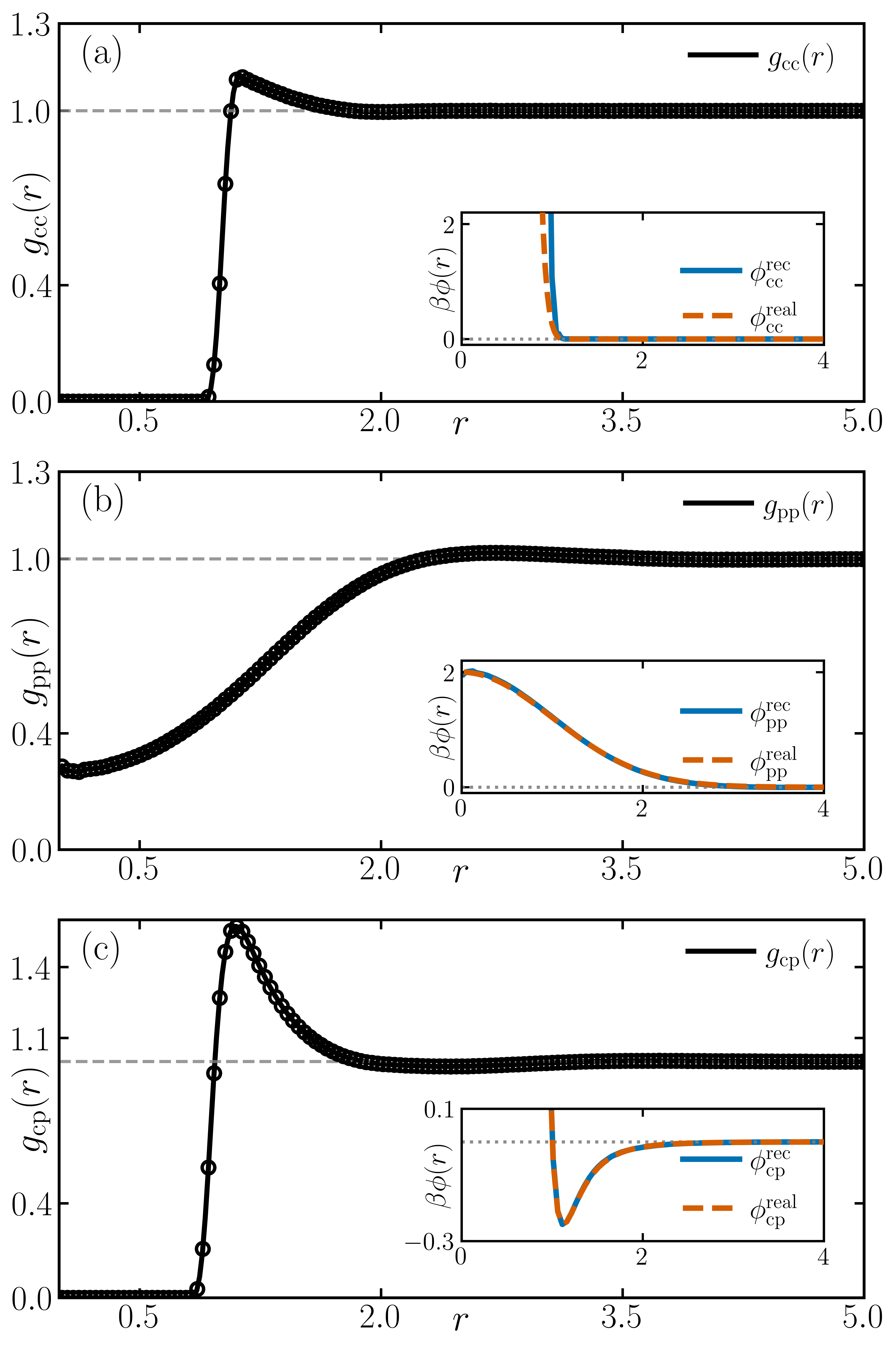}
\caption{
Structural correlations and reconstructed pair potentials for the colloid--polymer system. Panels (a)--(c) show the radial distribution functions $g_{\rm cc}(r)$, $g_{\rm pp}(r)$, and $g_{\rm cp}(r)$ obtained from simulations. The colloid--colloid and polymer--polymer correlations are calculated from one-component systems containing only the corresponding species. The insets compare the interaction potentials used in the simulations (dashed lines) with those reconstructed by IBI (solid lines). The reconstruction is performed at $\sigma^3\rho_{\rm c}=0.01$ and $\sigma^3\rho_{\rm p}=0.54$.
}
\label{fig::inverse_boltzmann}
\end{figure}

As shown in Fig.~\ref{fig::inverse_boltzmann}, the reconstructed potentials closely reproduce the corresponding model potentials for all three interaction pairs. The purpose of this reconstruction, however, is not to recover the underlying interaction potential as an end in itself. Rather, it provides a state-point-specific decomposition from which the repulsive reference contribution and the corresponding effective particle size can be determined.

For interactions containing an excluded-volume core, the reconstructed potential is subsequently decomposed using the WCA scheme,
\begin{equation}
\phi(r)=\phi^{\rm ref}(r)+\phi^{\rm tail}(r),
\end{equation}
where
\begin{equation}
\phi^{\rm ref}(r)=
\begin{cases}
\phi(r)-\phi(R), & r<R,\\
0, & r\ge R,
\end{cases}
\label{eq:wca_reference}
\end{equation}
and
\begin{equation}
\phi^{\rm tail}(r)=
\begin{cases}
\phi(R), & r<R,\\
\phi(r), & r\ge R.
\end{cases}
\label{eq:wca_attractive}
\end{equation}
Here, $R$ is the position of the minimum of the reconstructed potential. The reference contribution isolates the short-range repulsive core associated with excluded-volume effects, while the remaining contribution is treated as the tail. For the purely soft polymer--polymer interaction, no excluded-volume reference is required, and the reconstructed potential $\phi^{\rm p}(r)$ is retained directly in the soft-particle formulation.

The reference radial distribution function, $g^{\rm ref}(r)$, is then calculated for interactions containing an excluded-volume contribution and used to determine the optimized hard-sphere diameter, $\sigma_{\rm opt}$, by fitting the corresponding hard-sphere radial distribution function with $g^{\rm ref}(r)$. The hard-sphere potential is
\begin{equation}
\phi_{\rm hs}(r)=
\begin{cases}
\infty, & r<\sigma_{\rm opt},\\
0, & r\ge\sigma_{\rm opt},
\end{cases}
\label{eq:hard_core_potential}
\end{equation}
with the corresponding reference correlations obtained from the OZ equation using the PY closure.

The IBI reconstruction is therefore used only as an intermediate step to identify the repulsive reference contribution, determine the optimized particle size, and obtain the corresponding reference correlations. These quantities are then used to construct the free-energy functional directly from structural information, providing the connection between experimentally accessible correlations and thermodynamic properties developed in the following section.

\subsection{Integral equation theory for a one component system}
\label{sec:A2}

The structure of a homogeneous fluid is characterized by the radial distribution function $g(r)$, with the corresponding total correlation function defined as
\begin{equation}
h(r) = g(r) - 1.
\label{eq:correlation_and_rdf_1c}
\end{equation}
Experimentally, structural information is typically obtained through the static structure factor $S(k)$, which is related to the Fourier transform of the total correlation function via
\begin{equation}
S(k) = 1 + \rho \hat{h}(k),
\label{eq::structure_factor}
\end{equation}
where $\rho$ is the number density.

A formal connection between $h(r)$ and $c^{(2)}(r)$ is provided by the OZ equation,
\begin{equation}
h(r) = c^{(2)}(r) + \rho \int c^{(2)}(|\mathbf{r}-\mathbf{r}'|)\, h(r')\, d\mathbf{r}',
\label{eq::oz_real_space_1c}
\end{equation}
which, in Fourier space, reduces to
\begin{equation}
\hat{h}(k) = \frac{\hat{c}(k)}{1 - \rho\, \hat{c}(k)}.
\label{eq:oz_solution_1c}
\end{equation}
This leads directly to
\begin{equation}
S(k) = \frac{1}{1 - \rho\, \hat{c}(k)},
\label{eq:structure_factor_cr}
\end{equation}
thereby establishing a direct link between experimentally measurable structural quantities and the DCF. This relation forms the basis for expressing thermodynamic properties in terms of structural correlations.

\subsection{Integral equation theory for a multi-component system}
\label{sec:A3}

In a multicomponent fluid, the structure is characterized by the set of partial radial distribution functions, $g_{ij}(r)$, and the corresponding total correlation functions,
\begin{equation}
h_{ij}(r)=g_{ij}(r)-1,
\end{equation}
where $i,j=1,\ldots,n$ label the components and $n$ is the total number of components in the system. Experimentally accessible partial structure factors are related to $h_{ij}(r)$ via
\begin{equation}
S_{ij}(k) = \delta_{ij} + \sqrt{\rho_i \rho_j}\,\hat{h}_{ij}(k),
\end{equation}
where $\rho_i$ is the number density of species $i$.

The multicomponent OZ equations couple all correlation functions through
\begin{equation}
h_{ij}(r) = c^{(2)}_{ij}(r) + \sum_k \rho_k \int c^{(2)}_{ik}(|\mathbf{r}-\mathbf{r}'|)\, h_{kj}(r')\, d\mathbf{r}',
\end{equation}
which can be expressed in Fourier space in a matrix form as
\begin{equation}
\hat{\mathbf{H}}(k) =
\left[\mathbf{I} - \hat{\mathbf{C}}(k)\,\boldsymbol{\rho}\right]^{-1}
\hat{\mathbf{C}}(k).
\end{equation}

To close the OZ hierarchy, an approximate closure relation is required. Within the hypernetted-chain (HNC) approximation \cite{hansen2013theory}, the closure generalizes to
\begin{equation}
c^{(2)}_{ij}(r) = \exp\big[-\beta \phi_{ij}(r) + \gamma_{ij}(r)\big] - \gamma_{ij}(r) - 1,
\end{equation}
where $\gamma_{ij}(r) = h_{ij}(r) - c_{ij}(r)$ denotes the indirect correlation function.
For systems with strong excluded-volume interactions, such as colloidal suspensions, the PY closure \cite{hansen2013theory} is often more appropriate and can be written as
\begin{equation}
c_{ij}(r) = \left[1 + \gamma_{ij}(r)\right]\left[e^{-\beta \phi_{ij}(r)} - 1\right].
\end{equation}

Several other closures are available, each with its own advantages and limitations. The choice and accuracy of a closure generally depends on the nature of the system and the interactions being considered \cite{pihlajamaa2024comparison}.

The numerical strategy for estimating effective particle sizes and reference structures follows the same procedure as in the one-component case.

\begin{figure}[h]
\includegraphics[width=\linewidth]{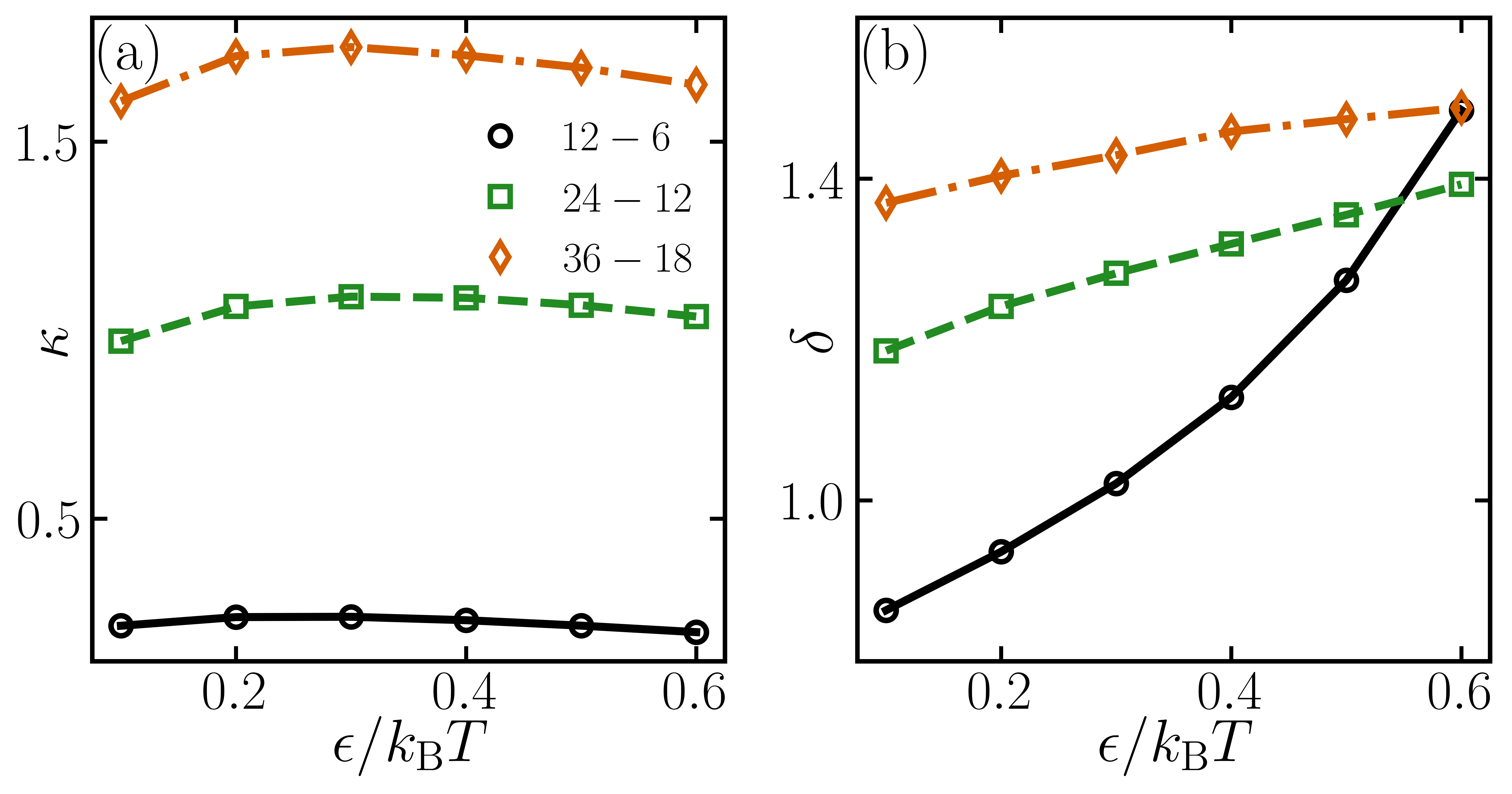}
\caption{
Fitting parameters (a) $\kappa$ and (b) $\delta$ are shown as a function of strength of interaction $\epsilon$, for different Mie potentials (see legend), with $\sigma = 1$.
}
\label{fig::rho_fitting}
\end{figure}

\subsection{Density dependence of the integrated strength}
\label{sec:A4}

To assess the behavior of the system near the critical point, we fit the integrated strength $A$ as a function of density using the two-parameter form
\begin{equation}
A(\rho) = \Delta B_2 \left(1 + \kappa \rho^{\delta} \right).
\label{eq::integrated_strength_curve_fitting}
\end{equation}
Here, the factor $\Delta B_2$ accounts for the dilute regime. The fitting parameters $\kappa$ and $\delta$ depend on the steepness of the potential, characterized by the exponents of the generalized Mie potential, as shown in Fig.~\ref{fig::rho_fitting}(a) and (b), respectively, for state points above the critical point. Both parameters exhibit an approximately monotonic variation with increasing interaction strength, as controlled by the attraction parameter $\epsilon$. This parametrization provides a means of extrapolating the density dependence of $A$ into the spinodal region, where direct integral-equation calculations and experimental measurements become increasingly difficult \cite{lutsko2007density}. The extrapolated behavior should nevertheless be regarded with caution, as it cannot be independently validated in this regime.

\bibliographystyle{apsrev4-1}
\bibliography{main}

\end{document}